%% file: SatComSPMagazine.tex
\pdfoutput=1
\documentclass[11pt, draftcls, onecolumn]{IEEEtran}
\usepackage
[
        a4paper,
        left=1.25in,
        right=1.25in,
        top=1in,
        bottom=1in,
]
{geometry}

\usepackage{array}

\usepackage{url}
\usepackage{color}
\usepackage{cite}

\usepackage{graphicx}
\usepackage[cmex10]{amsmath}
\usepackage{amsfonts}
\usepackage[tight,footnotesize]{subfigure}

\usepackage[]{algorithm2e}

\usepackage{amsmath}
\usepackage{amsfonts}
\usepackage{amssymb}

\begin{document}
%
\title{ Signal Processing for High Throughput Satellite Systems:
Challenges in New Interference-Limited Scenarios }

\author{Ana I. Perez-Neira$^{\star \dagger}$ Miguel Angel Vazquez$^{\star}$, Sina Maleki$^{\ast}$, M. R. Bhavani Shankar$^{\ast}$ and Symeon Chatzinotas$^{\ast}$,  
\\$^{\star}$ Centre Tecnol\`ogic de Telecomunicacions de Catalunya (CTTC/CERCA) 
\\$^{\dagger}$ Department of Signal Theory and Communications Universitat Polit\`ecnica de Catalunya \\
$^{\ast}$ SnT, Interdisciplinary Centre for Security, Reliability and Trust, University of Luxembourg \\
Email:\{\texttt{aperez}\}\texttt{@cttc.cat}
\\}


%


\maketitle

\begin{abstract}
The field of satellite communications is enjoying a renewed interest in the global telecom market, and very high throughput satellites (V/HTS), with their multiple spot-beams, are key for delivering the future rate demands. In this article the state-of-the-art and open research challenges of signal processing techniques for V/HTS systems are presented for the first time, with focus on novel approaches for efficient interference mitigation. The main signal processing topics for the ground, satellite, and user segment are addressed. Also, the critical components for the integration of satellite and terrestrial networks are studied, such as cognitive satellite systems and satellite-terrestrial backhaul for caching. All the reviewed techniques are essential in empowering satellite systems to support the increasing demands of the upcoming generation of communication networks. 
\end{abstract}


%
\IEEEpeerreviewmaketitle

\section{Introduction}

\input{introduction.tex}

\section{Precoding in Multibeam Satellite Systems}
\label{Precoding in Multibeam Satellite Systems}

\input{precoding.tex}

\section{User Terminal-Guided Non-Orthogonal Access}

\input{mud.tex}

\section{Onboard Signal Processing}
\input{obp.tex}
\section{Flexible Communications and Hybrid Solutions}
\input{flexible.tex}

\section{Conclusions and Future Prospects}
\input{conclusions.tex}




\bibliographystyle{IEEEtran}

\bibliography{IEEEabrvCTTC}

%
%
%

\end{document}

%% file: introduction.tex
In the past few decades, satellite communication (SatCom) systems have exploited new techniques and technologies that were originally implemented in terrestrial communications. For instance, while in the mid-1980s advanced analog-to-digital and digital-to-analog converters (ADC and DAC, respectively) were used in delay-sensitive audio/voice applications, satellite systems adapted them into more complex digital signal processing techniques in delay-tolerant video broadcasting \cite{Cor07}. Adaptation is critical due to the peculiarities of the SatCom system when compared to its terrestrial counterparts, including satellite channels, system constraints, and processing. 

Today there are approximately 1300 fully operational communication satellites. Every type of orbit has an important role to play in the overall communications system. Geostationary earth orbit (GEO), at 35,000 km, present an end-to-end propagation delay of 250 ms; therefore, they are suitable for the transmission of delay-tolerant data. Medium earth orbit (MEO), at 10,000 km, introduce a typical delay of 90 ms; based on that, they can offer a compromise in latency and provide fiber-like data rates. Finally, low earth orbit (LEO) is at between 350 and 1,200 km, and introduce short delays that range from 20 to 25 ms. In all these cases, the satellite is a very particular wireless relaying node, whose specificities lead to a communication system that cannot be treated like a wireless terrestrial one. This is because the channel, communication protocols, and complexity constraints of the satellite system create unique set of features \cite{booksat}, notably:

\begin{itemize}
\item Due to the long distance to be covered from the on-ground station to the satellite, the satellite communication link may introduce both a high round-trip delay and a strong path-loss of hundreds of dB. To counteract the latter, satellites are equipped with high-power amplifiers (HPA) that may operate close to saturation and create intermodulation and non‐linear impairments. 
\item Satellite communications traverse about 20 km of atmosphere and introduce high molecular absorption, which is even higher in the presence of rain and clouds, particularly for frequencies above 10 GHz. Therefore, satellite links are designed based on thermal noise limitations and on link budget analysis that considers large protection margins for additional losses (e.g., rain attenuation).
\item In the non-geostationary orbits (i.e., MEO and LEO), there are high time-channel variations due to the relative movement of the satellites with respect to the ground station.
\item Due to the long distance and carrier frequencies, the satellite antenna feeds are generally seen as a point in the far-field, thus making the use of spatial diversity schemes challenging. Also, due to the absence of scatters near the satellite (i.e., there are no objects in space that create multiple paths) and the strong path-loss (i.e., it is a long-distance communication), the presence of a line-of-sight component, which focuses all the transmitted power and is not blocked or shadowed, is much more critical than in terrestrial cellular communications. On the positive side, due to the lack of rich scatters, satellite communications experience higher cross-polarization isolation than terrestrial communication networks.
\item The processing complexity on-board the satellite is limited, as it is highly correlated with its power consumption, mass, and ultimately, with the final cost of the system. 
\item The received signal-to-noise Ratio (SNR) is very low and therefore the user terminal (UT) must have high sensitivity, good receiver antenna gains, and good tracking capabilities to steer the beam of the UT such that it continuously points to the satellite.
\item The practical challenges of the satellite system require solutions that are different from the ones used in the terrestrial wireless communications. An important one is the specific satellite multi-user protocol framing that is defined in the current broadcast and broadband standards (i.e., DVB-S2X). In these protocols, in order to overcome the satellite channel noise, channel codes are long and, therefore, must take into account data from multiple users. This fact creates a multicast transmission, because the same information has to be decoded by a group of users. Multicast transmission creates specific precoding techniques, as section II explains.
\item Finally, satellite solutions are generally characterized by a relatively long development phase before deployment. This is different from terrestrial solutions, where it is easier to test new technologies \textit{in situ} without incurring in excessive deployment costs. 
\end{itemize}

In the past few years, two important new trends have been observed in the satellite sector. The first one relies on the vast potential of the new generation of the so-called very high and high throughput satellite (V/HTS), as is explained in the next sub-section. Many operators are currently upgrading their constellations to deliver higher radio frequency (RF) power, enhanced functionality, and higher frequency reuse with V/HTS technology. The second one takes into account the fact that terrestrial wireless communications are going up in frequency and, due to that, the coexistence with the SatCom systems for using the same frequency bands will be needed. These new trends pose interesting challenges regarding new interference-limited scenarios, and signal processing (SP) offers valuable tools to cope with them. Before going more into the details of these new challenges, let us comment about the actual and future context of SatCom services.

Satellite communications have specific advantages with respect to terrestrial communications. For example, SatCom systems provide ubiquitous coverage. Currently, satellite systems, supported by their inherent wide coverage, are considered essential in satisfying the increasing data traffic ubiquity, which is expected to continue to increase over the coming decades. Satellites are capable of addressing wide geographic regions, even continents, using a minimum amount of infrastructure on the ground. The second feature stems from the broadcast nature of the satellite, which facilitates the delivery of the same content to a very large number of users. The ubiquitous coverage, together with the efficiency of its broadcast nature, improves the area data traffic and the communications mobility that can be supported. SatCom is the only readily available technology capable of providing connectivity anywhere, regardless the end user is fixed or in a moving platform on the ground, sea or air (e.g., on a train, ship or airplane). Finally, energy efficiency is also another key advantage, in which SatCom can play an important role in the need to reduce energy demands. This is because, once the satellite is in space, it has access to solar energy and can stay in orbit for up to 15 years with no real estate costs. Thanks to these features, the SatCom ecosystem on its own is efficiently serving very specific private and public sectors, for example, resilient overlay communications and disaster relief, governmental services, traffic off-loading and remote cellular backhaul provisioning, multicast services, and SCADA (supervisory control and data acquisition) for tele-supervision of industrial processes. While such a diversification of satellite-only services is foreseen to bear fruit, maximum benefits are envisaged by integrating satellite and terrestrial communications in the future fifth generation (5G) communications. Potential new markets and emerging applications that are currently pursued by the satellite community include ubiquitous broadband access, commercial aeronautical and maritime services, machine-to-machine communications, and smart cache feeding. In all these applications, SP is challenged to satisfy the corresponding requirements in terms of spectrum and energy. To sum up, the increase in demand for these new satellite services and systems is driving innovative approaches that are moving away from the traditional linear television broadcast (i.e., direct to the home, or DTH). Let us now introduce V/HTS, which is a key technology in this paradigm shift.

\subsection{High Throughput Satellites: A New Interference-Limited Paradigm}

\begin{figure}[h!]
	\centering
    \includegraphics[width=0.8\textwidth]{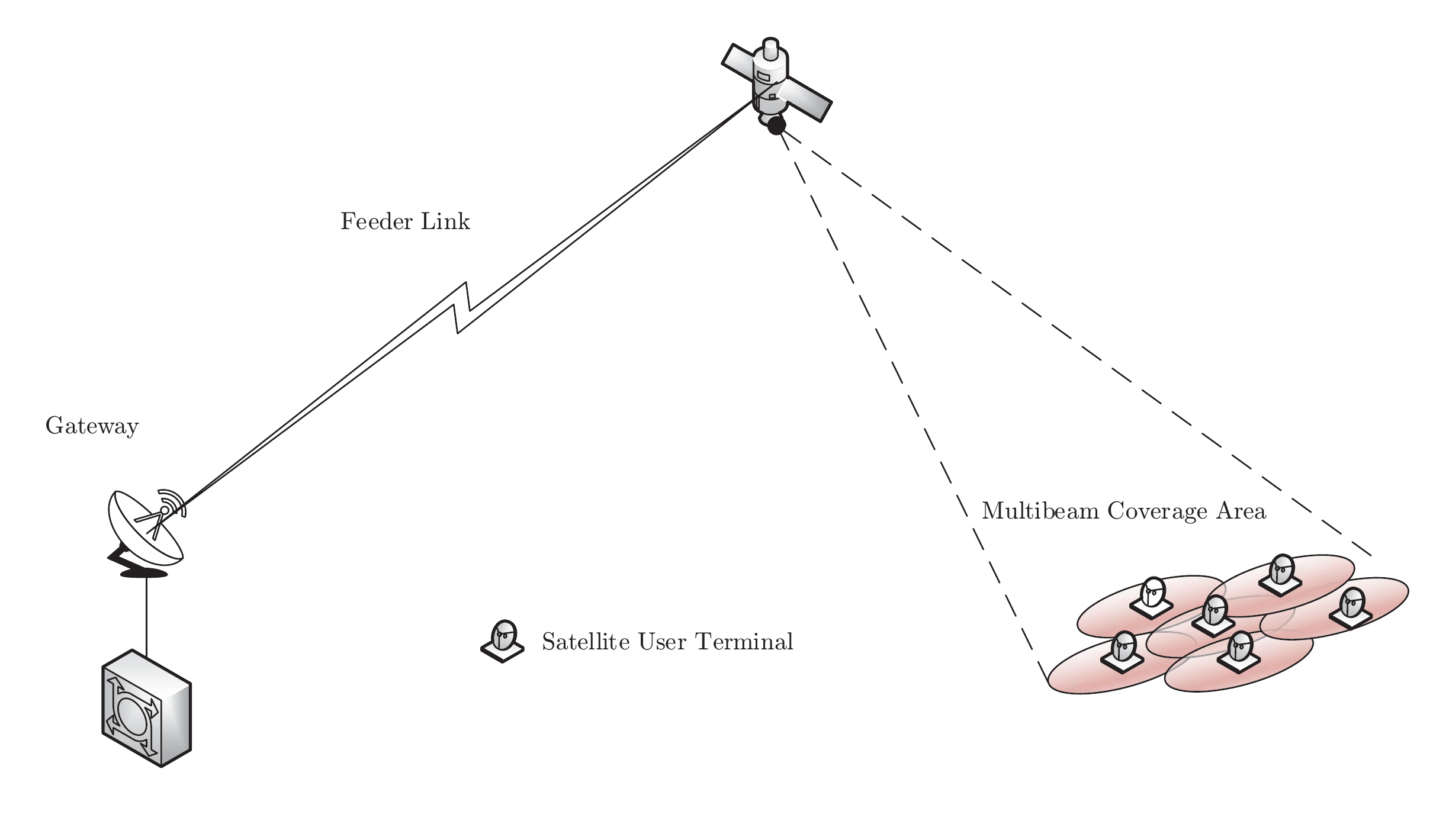}
    \centering
		\caption{Scheme of the multibeam satellite system. The forward link goes from the GW to the UTs via de satellite. The reverse link goes from the UTs to the GW via the satellite, too.}
      \label{arch}
\end{figure}

In contrast to mono-beam satellites, high throughput satellites split the service area into multi-spot beam service areas, which allows higher aggregate throughput and more service flexibility to satisfy a heterogeneous demand. The system architecture is shown in Fig. \ref{arch} and comprises a Gateway (GW), a satellite, and multiple UTs. The gateway (GW) is connected to the core network and serves a set of users that are geographically far away using the satellite as relaying node. The link from the GW to the satellite, and from the satellite to the UT are known as the feeder link and the user link, respectively. In the usual star configuration that is observed in Fig. \ref{arch}, the feeder link presents high directivity and gain. As this link presents a SNR that is considerably higher than the one in the user link, it is assumed in general to be noiseless and perfectly calibrated against channel power variations due to atmospheric events. Also, depending on the direction of the communication, the link receives the name forward link when it goes from the GW to the UT and reverse link when it goes from the UT to the GW. Each of the four mentioned links usually works in a different frequency band. The frequency selection is driven by many considerations, among them coverage and beam size, atmospheric conditions in the served region, and availability of a robust ecosystem of ground equipment technologies. For instance, current-generation GEO HTSs typically use the Ka-band, which is less congested than the C/Ku-band. For fixed satellite services (FSS), this refers to the exclusive satellite band from 19.7 to 21.2 GHz for the forward link and from 29.5 to 31 GHz for the reverse link. In land mobile satellite services (MSS) generally use lower frequencies such as the L-band (i.e., from 1.5 to 2.5 GHz) because of its lower attenuation, which enables a less complex UT. Note, however, that recently the Ka-band is also being considered to provide in-flight and maritime connectivity.

The HTSs that are currently operative (e.g., Viasat-2, SES-12) provide aggregate data rates of more than 100 Gbps. These HTS systems use the Ku/Ka-band in both feeder and user link, and serve in the user link as much as 200 beams in the same frequency band. VHTS systems (e.g., Viasat-3) aim at achieving data rates in the range of Tbps and, due to that, they need higher frequencies in the Q-band ($30-50$ GHz), V-band ($50-75$ GHz), and W-band ($75-110$ GHz), in order to serve as much as 3000 beams in the user link. For these reasons, advanced SP is required in order to reduce the interference among so many multiple beams, facilitate adaptive coverage, dynamically optimize the traffic, and share the spectrum with terrestrial services, among other functions. Flexibility in the resource allocation per beam can significantly improve the quality of service and bring down the incurred cost of the V/HTS system per transmitted bit.   

\begin{figure}[h!]
	\centering
    \includegraphics[scale = 0.5]{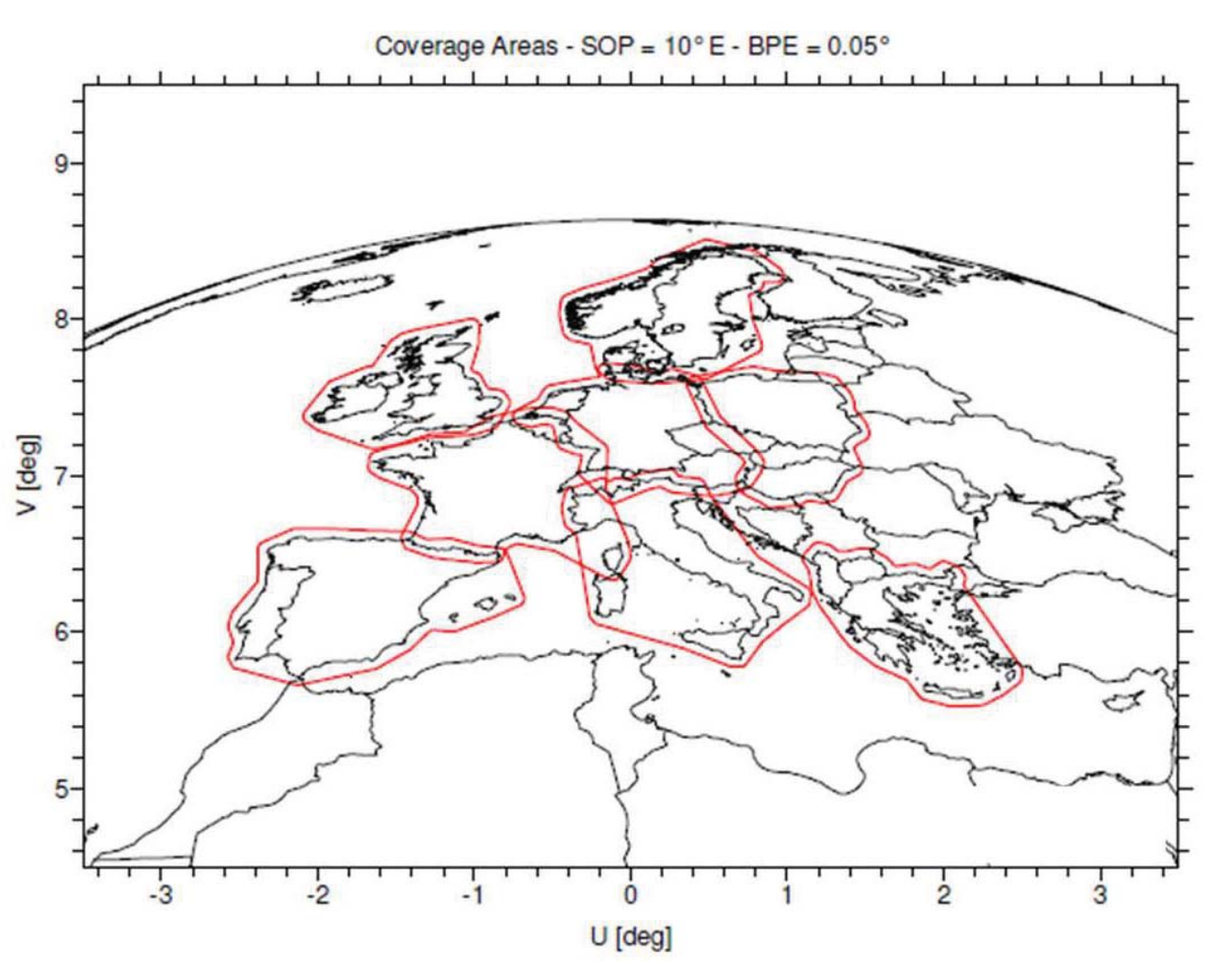}
    	\centering
		\caption{Broadcasting satellite with eight linguistic beams in the Ka-band (copyright European Telecommunications Standards Institute, 2015; further use, modification, copy and/or distribution are strictly prohibited).}
      \label{linguistic}
\end{figure}

\begin{figure}[h!]
	\centering
    \includegraphics[scale = 0.5]{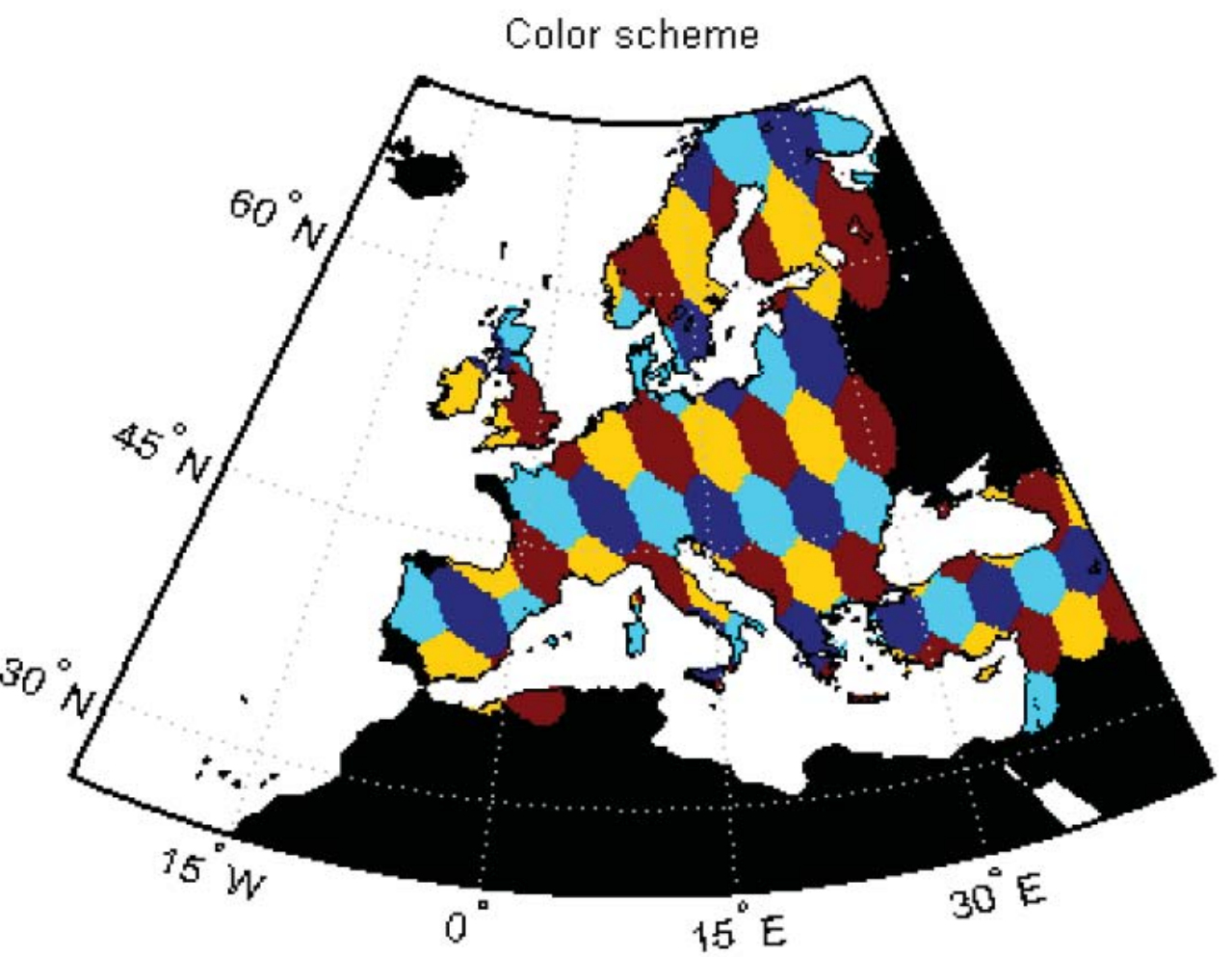}
     \centering
		\caption{User frequency plans for the scenario with 71 beams and frequency re-use of four (copyright European Telecommunications Standards Institute, 2015; further use, modification, copy and/or distribution are strictly prohibited).}
      \label{color}
\end{figure}

Fig. \ref{linguistic} shows an example of the classical linguistic beam wide coverage. In contrast, multi-spot beams allow tessellation of the coverage into much smaller footprints, thus enabling frequency reuse within the geographical area covered by one linguistic beam. As a consequence, per user bandwidth assignment and the aggregate throughput can potentially increase in V/HTS. Multi-spot beams enable broadband data services in addition to the traditional broadcast services offered by the linguistic beams. Fig. \ref{color} shows an example of the footprints of a four-color reuse scheme, where a total bandwidth of 500 MHz is allocated to the user link at the Ka-band. This bandwidth is divided into two sub-bands that, when combined with two orthogonal polarizations, generates the so-called four-color beam pattern across the coverage area. In the Ku/Ka-band, orthogonal polarizations maintain very low cross-polarization and due to that, they can be used as if they were different frequencies. Within each beam, multiple users are served with time division multiple access (TDMA). Currently, with the common frequency reuse of four colors, the interference power among beams is in the range from 14 to 34 dB below the carrier signal.

With the aim of lowering the cost per transmitted bit and increasing the spectral efficiency or the available system bandwidth, new systems aim at reusing more aggressively the available spectrum among the spot beams. Nevertheless, increasing the frequency reuse leads to a further increase of intra-system interference among the co-channel beams, which shifts the classical noise-limited link budget analysis towards an interference-dominated situation. The sidelobes of the beam radiation patterns create interference leakage among beams, and the carrier-to-interference ratio (CIR) can be severely degraded. In order to successfully implement high frequency reuse, interference management has to be implemented at the gateway, the satellite or the UT or some combination of these. It follows that the CIR mostly depends on the position of the UT, the cross-over level, and the antenna radiation pattern. Hence, the most favorable case corresponds to the situation in which the UT is in the center of the beam, while the worst case is when the UT is located at the beam-edge area. We note that for a frequency reuse pattern equal to one (i.e., $f_r = 1$) the average CIR in dB is around 0 dBs, for $f_r = 2$ it is 8 dB, for $f_r = 3$ it is 25 dB, and for $f_r = 4$ it is 30 dB. The interference power that comes from the high frequency reuse adds to that originating from the nonlinear distortion of the HPA. Unfortunately, the traditional approach to diminishing interference by using power control is insufficient, and, therefore, novel signal processing alternatives that exploit the structure of the co-channel interference structure are needed.

We note that the final V/HTS system performance depends not only on the capabilities of the applied signal processing, but also on many system choices. Complex design trade-offs and practical aspects need to be respected, as detailed in references like \cite{fenech12}. For example, if hundreds of beams are available in the system, high frequency reuse schemes can stress the payload resources of the satellite in terms of mass, power, and thermal dissipation. Another important consequence of increasing the frequency reuse is that the frequency bandwidth of the feeder link should increase accordingly. As this is not straightforward to do, different alternatives should be studied, such as employing multiple gateways in the feeder link (e.g., \cite{Joroughi2017-2}).

Finally, it is important to note that V/HTS systems require the most advanced transmission standards. Currently, DVB S2/S2X are the standards of both forward broadcast and broadband satellite networks. Using high efficiency modulation and coding schemes (MODCODs) up-to 256APSK combined with advanced interference management techniques enable aggressive and flexible frequency reuse. DVB-S2X incorporates the novel super-framing structure that enables the use of SP techniques that have never been used before in the satellite context, such as precoding and multi-user detection at the user terminal. Among other things, it incorporates orthogonal Walsh-Hadamard (WH) sequences as reference/training sequences, allowing simultaneous estimation of the channel state information of multiple beams. The super-frame concept was designed to maximize the efficiency of the channel coding scheme by encapsulating the information intended to several UTs using the same MODCOD. Remarkably, the length of the super-frame remains unaffected by the various transmission parameters that are applied on the different beams (e.g. MODCODs). Further details on DVBS2/S2X standard, which have a beneficial impact on precoding and multi-user detection, can be found in Annex E of \cite{DVB-S2X}. 

\subsection{Challenges and Organization of the Paper}

In the rest of the paper, we address the different SP techniques that we have identified as potential candiates to improve the data rate of future V/HTS systems. For each SP technique, we also mention the key implementation challenges that we have detected along with a possible solution that we have identified. The rest of the paper is organized as follows:
\begin{itemize}
\item Section II deals with spatial precoding techniques at the GW in order to mitigate the inter-beam interference. Note that a single V/HTS manages hundreds of feeds and controls a wide geographical area with a large number of users that typically have different traffic and quality of service (QoS) requirements. Thus, SP has to be studied for large-scale optimization in multibeam and multiuser systems. Due to the harsh interference among beams, these optimization problems are non-convex. 
\item Section III presents user terminal-guided SP. Spatial precoding requires channel state information at the transmitter (CSIT). However, if either partial or no CSIT is available, the system should resort to multi-user detection (MUD) capabilities at the UT in order to diminish interference. This section sets the framework and system model in order to devise and compare possible transmission schemes that incorporate receivers with MUD capabilities. 
\item Section IV deals with onboard processing (OBP) in the satellite, which introduces additional degrees of processing and performance improvement when compared to the traditional satellite approach that applies/uses a transparent payload. As expected, the ability to place OBP will dramatically change the integration of the satellites into the terrestrial networks.
\item Section V presents flexible communications and hybrid/integrated solutions. In this section, we discuss the spectral coexistance mechanisms through cognitive satellite communications, as well as integrated satellite-terrestrial backhauling for caching. In both cases, we discuss the techniques enabling these advances, as well as underlying problems that need to be solved using advanced SP techniques.  
\item Finally, section VI concludes the paper and discusses open lines of further research on this topic within the satellite community. 
\end{itemize}
This feature article provides for the first time an overview of the current state-of-the-art of the signal processing techniques, future perspectives, and challenges within the interference-limited scenarios that are emerging in V/HTS systems. The main topics are selected and structured. Instead of aiming at a broad-brush overview of the different satellite orbits and services, this paper focuses on the GEO FSS in the C/Ku/Ka-bands, where signal processing is needed to attain the promised Tbps rates.  It is also in the geostationary orbit where V/HTS has originated with well-established waveforms, coding, and modulators defined in the DVB-S2X standard. The use of non-GEO satellites (i.e., LEO and MEO) and MSS are discussed in the last section of this paper as open topics. Non-GEO V/HTS and mobile services still present many open questions from the signal processing point of view, due to the impact of the high-speed satellite movement that creates high Doppler spread, and time-varying gains.

%% file: precoding.tex
\subsection{Architecture and Communication Peculiarities}

With the aim of increasing the offered data rates of a given satellite, both operators and manufacturers are investigating a variety of alternatives. One main approach is to consider satellite communication links at extremely high frequencies such as the W-band \cite{Riva2014}. However, large investments are required for implementing the communication subsystems in these bands; in addition, new challenging channel impairments appear. As a result, spectrally efficient alternatives that exploit the current frequency bands are of great interest.

This is the case of precoding techniques that allow a high frequency reuse factor among different beams. With the aid of precoding, a satellite UT can obtain a sufficiently large signal to interference and noise ratio (SINR) even though the carrier bandwidth is reused by adjacent beams. In order to maintain a certain SINR value, the precoder mitigates the interference that can affect the  satellite UT.

Resorting to the system architecture depicted schematically in Fig. \ref{arch}, the precoding matrix is computed at the satellite GW. After that, the beam signals are precoded and transmitted through the feeder link using a Frequency Division Multiplexing (FDM) scheme. Then, the satellite payload performs a frequency shift and routes the resulting radio signal over an array-fed reflector antenna that transmits the precoded data over a larger geographical area that is served by the multiple beams in the user link.

\begin{figure}[h!]
	\centering
    \includegraphics[scale = 0.5]{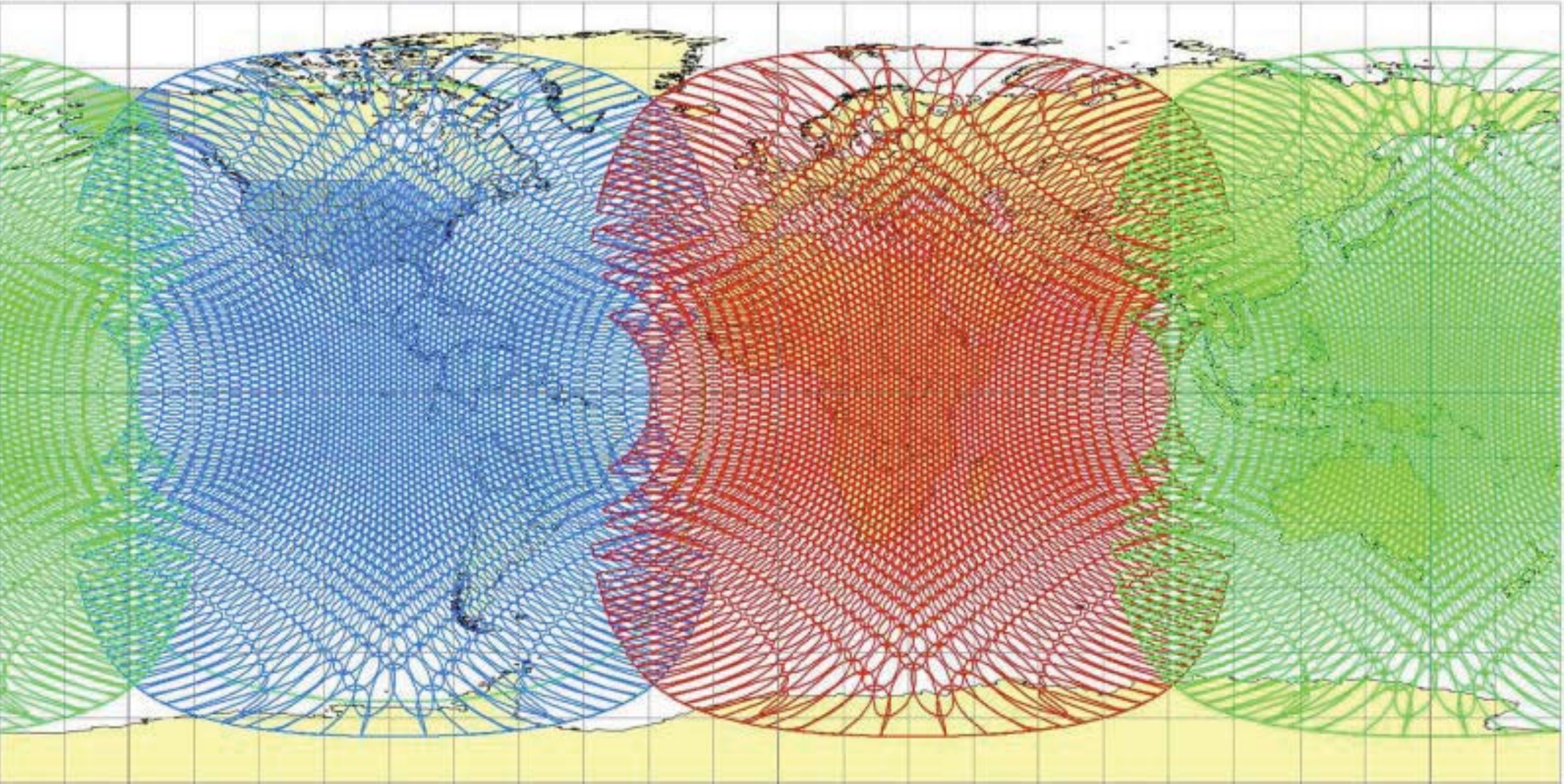}
    \centering
		\caption{Viasat 3 beampattern footprints. Each of the colors corresponds to three different satellite coverage areas. There are 1000 spots per color, which complicates the precoding implementation due to the extremely large size of the precoding matrix that must be calculated. (source: Viasat).}
      \label{viasat3}
\end{figure}

Multibeam precoded satellite systems can be modeled as a multiple-input-multiple-output (MIMO) broadcast channel \cite{Costa}. As it happens, in terrestrial systems, low complexity linear precoding techniques are of great interest. Indeed, the computational complexity that is required to implement multibeam satellite precoding techniques gains importance as the dimensions of multibeam satellite systems grow. For instance, the forthcoming Viasat-3 system is expected to utilize nearly 1000 beams to serve the coverage area that is presented in Fig. \ref{viasat3}. As a result, the on-ground equipment should be prepared to update a precoding matrix of 1000 users on a per-frame basis.

Bearing in mind the large dimensions of the multibeam satellite systems, the answer to the following question becomes crucial: \textbf{Is a multibeam satellite a massive MIMO system?} The short answer is \emph{no}, and is based on the following reasons:

\begin{enumerate}

\item \emph{The co-channel interference power does not decrease as the number of beams increases.} The favorable propagation in massive MIMO mentioned in \cite{massive1} does not occur in multibeam satellite systems. That is, in a scattered terrestrial channel environment, the off-diagonal elements of the channel covariance matrix tend to zero as the number of antennas grows, leading to an ideal interference-free scenario. On the contrary, due to the low scatter in the satellite channel, there is always strong co-channel interference among beams independent of the dimension of the multibeam satellite system. 

\item \emph{There is no pilot contamination.} Massive MIMO in multicell scenarios entails difficulties in the channel estimation operation as users located in adjacent cells might inject interference into the estimation process. In the multibeam satellite case, this does not occur since the number of adjacent beams in a given area is limited. Also, the pilot signals of adjacent beams are orthogonal, and the satellite channel is, in general, non-frequency-selective and preserves the orthogonality at the UT.

\item \emph{Multibeam satellite systems can naturally perform multicast transmission.} Due to the large coverage area of each satellite beam, which is in the order of few hundred kms, most of the satellite communication standards assume that a transmitted codeword would contain information from more than one UT, leading to a channel coding gain with respect to the case where short individual codewords are used.

\end{enumerate}

\subsection{Precoding Techniques}

Let us consider a multibeam satellite system in which the satellite is equipped with an array-fed reflector antenna with a total number of feeds equal to $N$. These feed signals are combined to generate a beam radiation pattern composed of $K$ beams, which is considered fixed. For each frame, we assume that a total number of $N_u$ users are simultaneously served per beam (i.e., the total number of served users by the satellite is $KN_u$). Considering that all beams radiate in the same frequency band (i.e., $f_{r}=1$), the instantaneous received signal at the $i$-th user terminal of each beam is given by
\begin{equation}\label{sys}
\mathbf{y}^{[i]} = \mathbf{H}^{[i]} \mathbf{x} + \mathbf{n}^{[i]}, \quad  i = 1, \ldots , N_u,
\end{equation}
where vector $\mathbf{y}^{[i]} \in \mathbb{C}^{K \times 1}$ is the vector containing the received signals of the $i$-th UT (i.e., the value $\left[\mathbf{y}^{[i]} \right]_k$ refers to the received signal of the $i$-th UT at the $k$-th beam), whereas vector $\mathbf{n}^{[i]} \in \mathbb{C}^{K \times 1}$ contains the noise terms of each $i$-th UT. The entries of $\mathbf{n}^{[i]}$ are assumed  to be independent and Gaussian distributed with zero mean and unit variance (i.e., $E\left[\mathbf{n}^{[i]} {\mathbf{n}^{[i]}}^H \right] = \mathbf{I}_{K} \quad i = 1, \ldots , N_u$). Finally, vector $\mathbf{x} \in \mathbb{C}^{K \times 1}$ contains all the transmitted signals.

The channel matrix can be described as follows:
\begin{equation}\label{cc}
\mathbf{H}^{[i]} = \mathbf{F}^{[i]} \circ \overline{\mathbf{H}}^{[i]}, \quad  i = 1, \ldots , N_u,
\end{equation}
where the $(k,n)$-th entry of matrix $\overline{\mathbf{H}}^{[i]}  \in  \mathbb{R}^{K \times N}$ is 

\begin{equation}\label{ccana}
\left[\overline{\mathbf{H}}^{[i]}  \right]_{k,n} = \frac{G_Ra_{kn}^{[i]} e^{j\psi_{k,n}^{[i]}}}{4 \pi \frac{d_{k}^{[i]}}{\lambda} \sqrt{K_BT_RB_W}}\;k=1, \ldots , K; n=1, \ldots , N; i=1, \ldots , N_u.
\end{equation}

$d_{k}^{[i]}$ is the distance between the $i$-th UT at the $k$-th beam and the satellite. $\lambda$ is the carrier wavelength, $K_B$ is the Boltzmann constant, $B_W$ is the carrier bandwidth, $G_R^2$ is the UT receive antenna gain, and $T_R$ is the receiver noise temperature. The term $a_{kn}^{[i]}$ refers to the gain from the $n$-th feed to the $i$-th user at the $k$-th beam. The time varying phase due to beam radiation pattern and the radio wave propagation is represented by $\psi_{k,n}^{[i]}$. Note that we have considered that the feeder link is ideal and its impact is limited to a scaling factor. 

Furthermore, matrix $\mathbf{F}^{[i]} \in \mathbb{C}^{K \times N}$ represents the atmospheric fading such that $\left[\overline{\mathbf{F}}^{[i]}  \right]_{k,n} = \mu_k^{[i]}e^{j\theta_k^{[i]}}$, where notably  each fading coefficient is independent of the transmission feed. That is, the UT experiences a fading value that equally impacts all feed signals. There is no multipath and a strong line-of-sight is present in frequencies above 10 GHz (i.e., above the Ku band), whenever there is no blockage.

In order to mitigate the co-channel interference due to the high frequency reuse factor, precoding is performed; therefore, the transmitted signal vector per beam is given by $\mathbf{x} = \mathbf{W}\mathbf{s},$ where $\mathbf{s} \in \mathbb{C}^{K \times 1}$ is the vector that contains the transmitted symbols per UT, which we assume are uncorrelated and unit normed $\left(E\left[\mathbf{s}\mathbf{s}^H \right] = \mathbf{I}_{K} \right)$. Matrix $\mathbf{W} \in \mathbb{C}^{N \times K}$ is the linear precoding matrix to be designed. As mentioned previously, each DVB-S2X frame contains information intended to multiple users in order to attain a large channel coding gain. In this context, every UT user with index $i=1,\ldots,N_{u}$ at the $k$-th beam shall detect the same information $[\mathbf{s}] _{k}$, leading to the so-called multigroup multicast transmission, which has already been studied for the general wireless systems in \cite{karipidis2008quality}.

The system sum-rate is defined as $\mathcal{SR} = \sum_{k = 1}^K \min_{i = 1, \ldots , N_u} \quad \log_2 \left(1 + \text{SINR}_k^{[i]} \right),$ where $\text{SINR}_k^{[i]}$ is the signal-to-noise-plus-interference ratio (SINR) of the $i$-th user at the $k$-th beam and is defined as
\begin{equation}\label{sinr}
\text{SINR}_k^{[i]}= \frac{|\mathbf{h}_k^{[i],H}\mathbf{w}_k|^2}{\sum_{j\neq k}|\mathbf{h}_k^{[i],H}\mathbf{w}_j|^2 + 1},
\end{equation}
where $\mathbf{h}_k^{[i],H}$ and $\mathbf{w}_k$ are the $k$-th row and $k$-th column of $\mathbf{H}^{[i]}$ and $\mathbf{W}$, respectively. Note that since we are considering a multicast transmission, the achievable data rate at each beam is determined by the data rate that the UT with the lowest SINR can achieve, as the selected MODCOD for transmission should be decodable by all UTs in the frame.

As a matter of fact, the system designer should find a solution to the following optimization problem:
\begin{equation}
\begin{aligned}
& \mathcal{P}_1: \quad  \underset{\mathbf{W} }{\text{maximize}} \quad  \mathcal{SR} \\
& \text{subject to}\\
& \left[\mathbf{W}\mathbf{W}^H \right]_{nn} \leq  P \quad n = 1 , \ldots , N. \\
\end{aligned}
\end{equation}
The optimization problem $\mathcal{P}_1$ is large-scale and non-convex (the objective function is non-convex). Note that in $\mathcal{P}_1$ a matrix of around 10,000 complex elements shall be optimized over hundreds of per-feed power constraints. The work in \cite{christopoulos2014weighted} considers the optimization of $\mathcal{P}_1$ via a semi-definite relaxation procedure, which is adequate from small to medium coverage areas. In case a notably larger number of beams and/or users are targeted, current non-convex optimization alternatives might fail due to the immense computational complexity; thus, opening potential avenues for future research.

A promising precoding alternative considering the performance-computational complexity trade-off is the 'UpConst Multicast MMSE' \cite{Taricco2014,Vazquez2016}, which can be written as $\mathbf{W}_{\text{MMSE}} = \beta_{\text{MMSE}}\left(\widehat{\mathbf{H}}^H\widehat{\mathbf{H}} + \frac{1}{P} \mathbf{I}_N \right)^{-1}\widehat{\mathbf{H}}^H,$ where $\beta_{\text{MMSE}}$ controls the transmit power to fulfill the per-feed power constraints and $\widehat{\mathbf{H}} = \frac{1}{N_u} \sum_{i = 1}^{N_u} \mathbf{H}^{[i]}.$ In other words, this design consists of minimum-mean-squared error (MMSE) precoding over the average channel matrix of all users simultaneously served at each beam. The channel elements in $\mathbf{H}^{[i]}$ are reported by each UT in the return link. Section III comments on the estimation of the UT’s channel. In a practical system the channel state information is usually quantized and contains residual errors, affecting the expected gains of the precoding techniques, when compared to the theoretical cases in which the CSIT is perfectly known. Fortunately, FSSs experience low satellite channel variability, and due to that, the existing studies report a performance improvement via satellite precoding. 

\begin{figure}[h!]
	\centering
    \includegraphics[scale = 0.5]{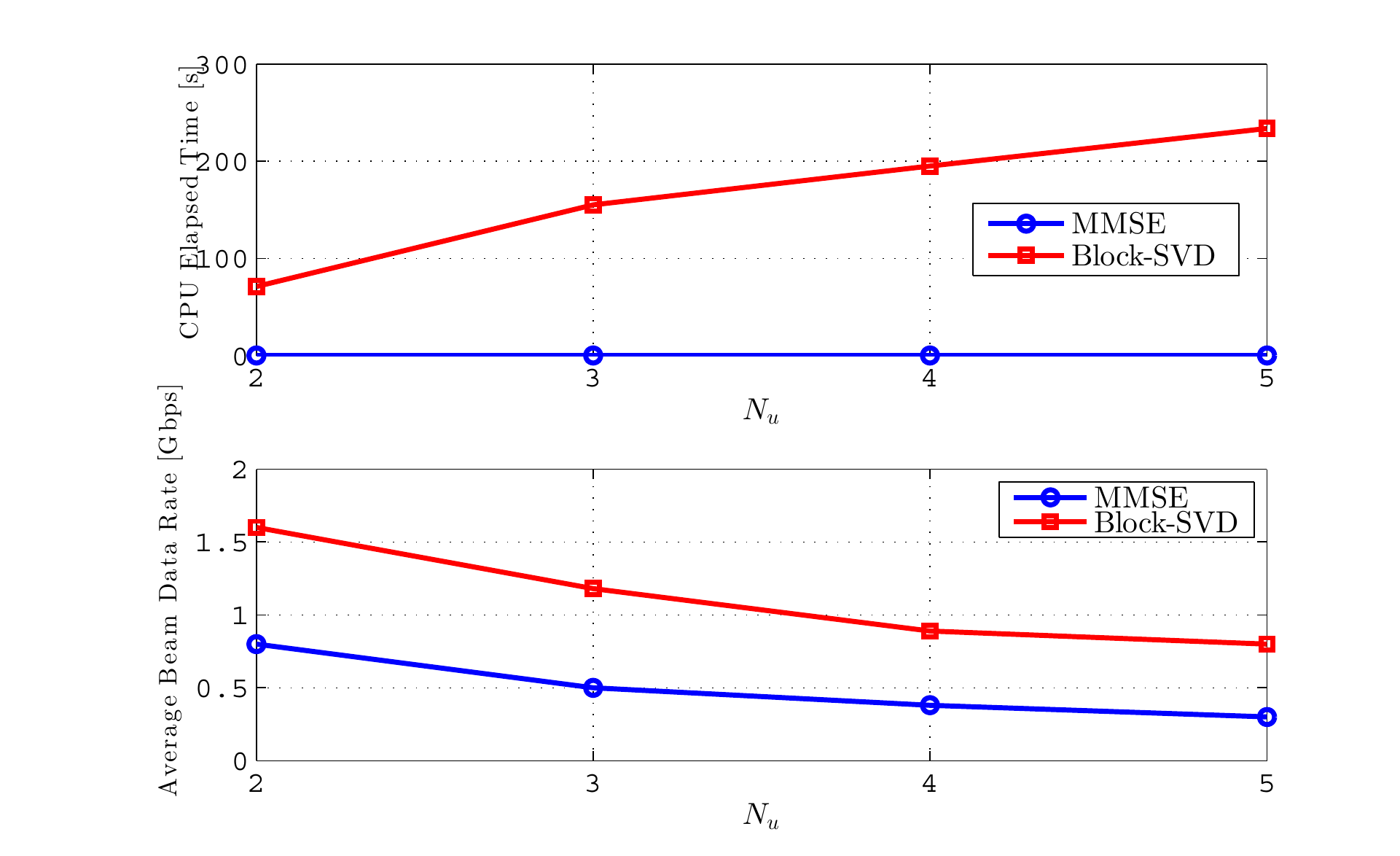}
    \centering
		\caption{Average beam data rate and average CPU elapsed time of two precoding techniques, MMSE and block-SVD.}
      \label{exam}
\end{figure}

A comprehensive study of linear precoding techniques for the general multigroup multicast communication model can be found in \cite{Silva2009}. Fig. \ref{exam} shows the beam data rate and the computational time of both 'upconst' multicast minimum mean square error (UpConst Multicast MMSE) and the block singular value decomposition (block-SVD) technique presented in \cite{Vazquez2016}. For obtaining the results, we consider a beampattern with 245 beams and a maximum per-feed power constraint of 55 W. Both the average central process unit (CPU) time and the average beam capacity, $SR/K$, have been obtained over 100 Monte Carlo runs. 

Clearly, the larger $N_u$, the lower are the attainable rates obtained by both block-SVD and UpConst Multicast MMSE. In all cases, Block-SVD leads to larger data rates compared to the UpConst Multicast MMSE. Nevertheless, the computational complexity of UpConst Multicast MMSE is much lower than Block-SVD and does not grow notably when the number of $N_u$ users per frame increases. On the contrary, block-SVD requires more computational time to compute the precoding matrix as the number of UT grows.

However, despite its low computational complexity,  UpConst Multicast MMSE still presents implementation challenges when serving large coverage areas (i.e., the computation of the matrix inverse becomes a computationally demanding operation as $K$ grows). Consequently, the study of alternative precoding designs is of extraordinary interest for both academia and industry. Concretely, it shall be explored precoding techniques that require a limited number of operations, when computing its precoding matrices, while they provide large data rates.

It is important to remark that the scheduling process plays a key role to obtain relevant sum-rate values; as it is crucial to select the most convenient users to be served in each satellite frame. As reported in \cite{Vazquez2016}, this scheduling process could just simply consider the geographical position of the UTs. In this way, information from UTs that are geographically close can be embedded into the same frame in order to yield efficient data rates. In any case, an open problem to be tackled is the characterization of the scheduling effect on the overall architecture, bearing in mind the queue's stability and the UT's targeted data rates.

Whenever the high layers are considered, the precoding design should be able to guarantee certain QoS to the UTs. In contrast to cellular systems, satellite operators offer their clients service level agreements (SLA) that involve a minimum data rate over a certain percentage of the channel access attempts. In this case, the fulfillment of the SLA contracts by precoding is done by optimizing the following problem:
\begin{equation*}
\begin{aligned}
& \mathcal{P}_2: \quad  \underset{\mathbf{W} }{\text{minimize}} \quad  \left||\mathbf{W} \right||^2 \\
& \text{subject to}\\
& \left[\mathbf{W}\mathbf{W}^H \right]_{nn} \leq  P \quad n = 1 , \ldots , N, \\
& \text{SINR}_k^{[i]} > \gamma_k \quad k = 1, \ldots , K \quad i = 1, \ldots , N_u.
\end{aligned}
\end{equation*}
The optimization problem $\mathcal{P}_2$ is a non-convex quadratically constrained quadratic problem (QCQP), which limits its applicability in large-scale coverage areas. This problem can be tackled via semi-definite relaxation (SDR) approximation methods such as the one in \cite{Karipidis2008}. Bearing this in mind, efficient parallel implementation of the non-convex QCQP optimization tools can be a good alternative for solving $\mathcal{P}_2$ in real multibeam satellite systems. This is the case for the work done in \cite{Vazquez2017}, which promotes the use of the consensus alternate direction of multipliers method of \cite{Huang2016} to solve the non-convex QCQP $\mathcal{P}_2$.

\subsection{Multiple Gateways}

Multibeam precoding over multiple GWs consists of transmitting the precoding signals over geographically separated GWs that are usually interconnected.  In this way the equivalent feeder link can aggregate the bandwidth of the feeder links of the different GWs and can accommodate the bandwidth increase that is needed when frequency reuse increases. In contrast to the single-gateway scheme, multiple-gateway precoding presents two main challenges.

First, the original precoding matrix $\mathbf{W}$ becomes block-diagonal so that
\begin{equation}\label{pmgw}
\mathbf{W} = \text{block-diag}\left\{ \mathbf{W}_1, \ldots , \mathbf{W}_l, \ldots , \mathbf{W}_L\right\},
\end{equation}
where $\mathbf{W}_l \in \mathbb{C}^{K_l \times N_l}$ is the precoding matrix associated with the $l$-th gateway $(l = 1 , \ldots , L)$. Note that for multiple-GW precoding $N = \sum_{l=1}^L N_l$, and $K = \sum_{l=1}^L K_l$. In other words, each GW can only use a subset of the $N$ feed signals for performing the interference mitigation. This fact limits the overall system performance as it reduces the available degrees of freedom. On the other hand, each of the GW feeder link bandwidth requirements is reduced. Indeed, the $l$-th gateway only transmits  $K_lN_l$ precoded signals instead of the $KN$ signals that were transmitted in the single-GW scenario.

The second main challenge is the channel state information acquisition. Each gateway can only access the feedback information from their served users, but each gateway needs the channel state information of the adjacent beams to reduce the generated interference. Therefore, a set of matrices must be exchanged by the different GWs, leading to a large communication overhead \cite{Joroughi2017-2}. Perfect connectivity between gateways might not be possible in real deployments. In this context, the multi-agent optimization of $\{\mathbf{W}_l \}_{l=1}^L$ may be of interest to implement assuming certain QoS requirements between the different GW connections. This impacts not only the tentative optimization, but also the design of the compression algorithms for exchanging information from the different GWs. Finally, the precoding structure in \eqref{pmgw} is similar to the group sparse beamforming. In light of this, promoting group-sparsity in both $\mathcal{P}_1$ and $\mathcal{P}_2$ might result in an efficient multi-GW precoding design.

%% file: mud.tex
The multibeam precoding techniques presented in Section II enable non-orthogonal multiple access, as it relies on CSIT at the GW. An alternative that relaxes the need for CSIT is to use multi-user detection (MUD) techniques at the UT. MUD can combat the inter-beam interference due to a high frequency reuse factor and lack of full CSIT. As the UT complexity is of paramount importance in keeping the cost of the overall satellite system data rate low, in this section we focus on UTs equipped with only one antenna, thus, no spatial interference rejection capability is possible. This section lays out a holistic comparative study using MUD at the UT, together with different non-orthogonal access strategies. Various possible satellite multibeam scenarios and CSIT requirements are taken into account. The obtained results are useful to identify some of the performance bounds of the different possible V/HTS access strategies.

As the complexity of MUD receivers grows exponentially with the number of signals to be detected, different simplification strategies for these detectors have been studied. It is not the aim of this section to review the MUD architectures or its simplifications. Instead, for interesting and useful designs we refer the reader to references like \cite{fer, bol}, where the sum-product algorithm or the single tree-search algorithm are studied, respectively. Although these schemes can achieve linear complexity in the number of interferers, in practice it is customary to limit the number of useful signals to two, or at most three, and treat the rest as background noise \cite{spaw}. For the sake of simplicity, we propose a MUD system model that limits the number of useful signals to two.

\subsection{System model}
For the sake of clarity, next we consider that two beams reuse the same frequency band and that in \eqref{sys} there is one user per beam (i.e., $N_{u}=1$). Although this setting is not compatible with current standards yet, it has been chosen because it is easy to explain, clear, and allows a presentation of broad scope that opens new avenues for research. These considerations yield the following two-user communication system:
\begin{equation}\label{mud1}
\begin{aligned}
& y_{1} = \mathbf{h}_{1}^{H}\mathbf{x}+z_{1} = h_{1}^{*[1]}x_{1}+h_{2}^{*[1]}x_{2}+z_{1}\\
& y_{2} = \mathbf{h}_{2}^{H}\mathbf{x}+z_{2}\ = h_{1}^{*[2]}x_{1}+h_{2}^{*[2]}x_{2}+z_{2},
\end{aligned}
\end{equation}
where the notation introduced in \eqref{sys} has been simplified as there is only one user per beam. Namely, $y_{i}$ for $i=1,2$ is the received signal in the beam $i$ and $x_{i}, i=1,2$ is the transmitted signal for the UT that has been served in that beam. Finally, $z_{i}$ (with equivalent noise power $\sigma_{i}^{2}$) combines the AWGN noise plus the residual interference term that is associated with the user in beam $i$. In a first approach, this paper considers perfect synchronization; in other words, symbol timing, carrier frequency, and phase can be estimated even under the challenging frequency reuse factor. In \cite{gap} the authors demonstrate that, under certain conditions, the modified Cramer-Rao lower bound for the mentioned synchronization parameters is the same both single- and multi-beam situation as in the multiple beam situation. These conditions basically allow a beam-wise decoupling of the estimation of the synchronism. To verify these conditions, the synchronization sequences must be orthogonal (e.g., as the Walsh-Hadamard sequences that are used in the DVB-S2X) and the gateway must pre-compensate the discrepancies that may appear in the time delays and frequency offsets among different transponders/beams. In case these conditions are not met and the signals received from different satellite beams are not perfectly synchronized in time and frequency, the UT will have to perform advanced frame, carrier, and timing synchronization.  The UT has to do these synchronization tasks as well as the estimation of the complex channel gains (i.e., amplitudes and phases). The authors of \cite{Pantelis2016} summarize these algorithms and show their performance in a multibeam scenario in presence of strong co-channel interference power for using a high frequency reuse factor.

Note that in the most general case, the signal $x_{i}$ that is fed into a given beam $i$ is a function of the symbols intended to several UTs. That is, the streams are not necessarily plainly multiplexed. Let $s_{k}^{[i]}$ be the signal that bears the message intended to the $i$-th user in beam $k$, or, equivalently, $s_{i}$ if there is only one user served per beam, then it follows that
\begin{equation}\label{stream}
\begin{aligned}
& x_{i} = f_{i}\left(s_{1},s_{2}\right) \quad i=1,2,
\end{aligned}
\end{equation}
where $f_{i}(.)$ can be any function. Without loss of generality, we assume in this section that $E\left[|x_{i}|^2\right]=P_{i}$, for $i=1,2$. 

\subsection{Achievable rates}

The communication system in \eqref{mud1} can be seen either an information-theoretic broadcast channel (BC) or an interference channel (IC). Therefore, by grouping adjacent beams in pairs, we can draw an analogy with the two-user BC or IC. These cases are definitely relevant because they have known close-to-optimal inner bounds on the capacity region.

If the beams cooperate and the power constraint is $E[|x_{1}|^2+|x_{2}|^2]=P$, then the BC model is the one to be considered. In the BC, different messages are simultaneously transmitted in the same frequency band, and each message is intended for a different receiver (note that this information theoretic concept of the BC differs from the concept of a broadcast service, where the same message is intended to all users that are in the coverage area. In the BC model, dirty paper coding, as proposed by M. Costa in \cite{Costa}, is the optimal strategy to achieve sum-capacity. It consists of optimally precoding the simultaneously transmitted signals while taking into account the interference that these signals are creating among each other in reception. Due to the interference, the transmission is done on a "dirty" environment, and this is where its name comes from. This optimal transmission requires full CSIT. However, when only partial or no CSIT is available, beams cannot cooperate and only independent power constraints can be considered: $x_{i}=\sqrt{P_{i}}$ for $s_{i}$ $i=1,2$. In this case, the IC model is the one to be considered.

The BC and IC are abstract channel models, and it is an open matter in the design of the satellite system when any of these two models is the most suitable to be developed for the specific multibeam satellite system of interest. For instance, if the goal is to obtain a low-complexity multi-user multibeam scheduler, the GW will work separately with each beam and with the users that lie within the coverage area of each of the beams; thus, the beams will not cooperate. As a consequence, those users that lie in the area where the beam footprints intersect will be managed by a hard hand-over. This strategy corresponds to an IC strategy and, although it reduces the performance of the system, the resulting complexity is low. The IC can also be the proper model when multiple GWs, which do not communicate among them, control the same satellite. In contrast, the BC is the one to be used whenever all the beams of the satellite are managed by a single GW. This GW has the complete CSIT of the system and can design fully multibeam cooperative precoders and user scheduling strategies.

The identification of an specific multibeam satellite system to design with the BC and IC abstract models has only been done recently, and it allows to use all the information theory bounds and SP techniques that are associated with any of the two channel models. The challenge is, however, how to implement, within the specific multibeam satellite system constraints (i.e., complexity, performance, cost), these available SP techniques or create more suitable ones when it is needed. As commented, whenever the UT has MUD capabilities, the GW can implement simple techniques with non/semi-cooperative beams. In other words, whenever the UT has MUD capabilities, the IC situation must be studied. Let us next revisit some existing strategies for the IC that can be suitable for the multibeam satellite. How the satellite system must be designed to comply with these strategies is open for further research.

The Han-Kobayashi (HK) inner bound is the best-known single-letter inner bound on the capacity region for the IC \cite{HK}. By using Gaussian codebooks simplified HK schemes reported in the literature that are demonstrated to be as close as $1$ bps/Hz from the capacity region. These simplified HK schemes do not need time-sharing and require that each code word is represented by a public and a private message, which are sent via superposition coding. This opens the floor to the so-called rate splitting approaches \cite{bru2}, whose implementation has to take into account that the public message is to be recovered by both receivers, whereas the private one has to be recovered only by its intended recipient. Rate splitting has come up as an interesting strategy when the transmission is overloaded (i.e., the number of simultaneous transmissions is greater than the number of feeds) or when the channel state information is non-existent or incomplete at the transmitter. Depending on the power that is allocated to the public and the private messages, different points within the capacity region can be reached. The different private messages create interference in the unintended receivers; the key benefit of rate splitting is to partially decode this interference and partially treat it as noise. This contrasts with the so-called non-orthogonal multiple access (NOMA) schemes, where the interference is treated either as noise or as useful signal by the UT. For instance, let us consider the IC situation with rate splitting. In this case, in \eqref{mud1} the transmitted signal for each user $i$ is generated by adding the public and the private signals, which are denoted by $x_{i1}$ and $x_{i2}$, respectively. That is, $x_{i}=x_{i1}+x_{i2}$, for $i=1,2$. To satisfy the power constraints, public and private codewords are subject to $E\left[|x_{i1}|^2\right]= (1-\lambda_{i})P_{i}, E\left[|x_{i2}|^2\right]= \lambda_{i}P_{i}$ for users $i=1,2$ and $0\leq \lambda_{i} \leq 1$.

The HK inner bound reduces to the \textbf{interference as noise} or \text{IAN} bound when $\lambda_{1}=\lambda_{2}=1$. In other words, it does not exploit the interference to improve the rate. Instead, if the transmission is asymmetric and, for instance, there is only a public message for user $1$ (i.e., $\lambda_{1} = 0$) and only a private message for user $2$ (i.e., $\lambda_{2} = 1$), then \textbf{sequential cancellation decoding} or \text{SCD} is more suitable. In other words, receiver $1$ treats the interference as noise (\text{IAN}) and receiver $2$ is able to recover the interfering signal by performing \text{SCD}. Analogously, the rates can be found if receiver $1$ and receiver $2$ exchange the decoding strategy (i.e., $\lambda_{1} = 1$, $\lambda_{2} = 0$).

 Finally, we comment on \textbf{simultaneous non-unique decoding} or \text{SND}. In \text{SND} each receiver $i$ tries to jointly decode $x_{i}$ and $x_{j}$ ($i,j=1,2$ with $i\neq j$), but user $i$ does not care about the errors when decoding $x_{j}, j \neq i$. In other words, if the modulation/coding assigned to user $j$ ($j \neq i$) is given beforehand, i.e., $M_{j}$, user $i$ does not decode $x_{j}, j \neq i$ if it is received with lower quality such that

\begin{equation}
M_{j}\geq \log_{2}\left(1+\frac{P_{j}|h_{i}^{[j]}|^2}{\sigma_{i}^{2}}\right).
\end{equation} 

\begin{figure}[h!]
	\centering
    \includegraphics[scale = 0.5]{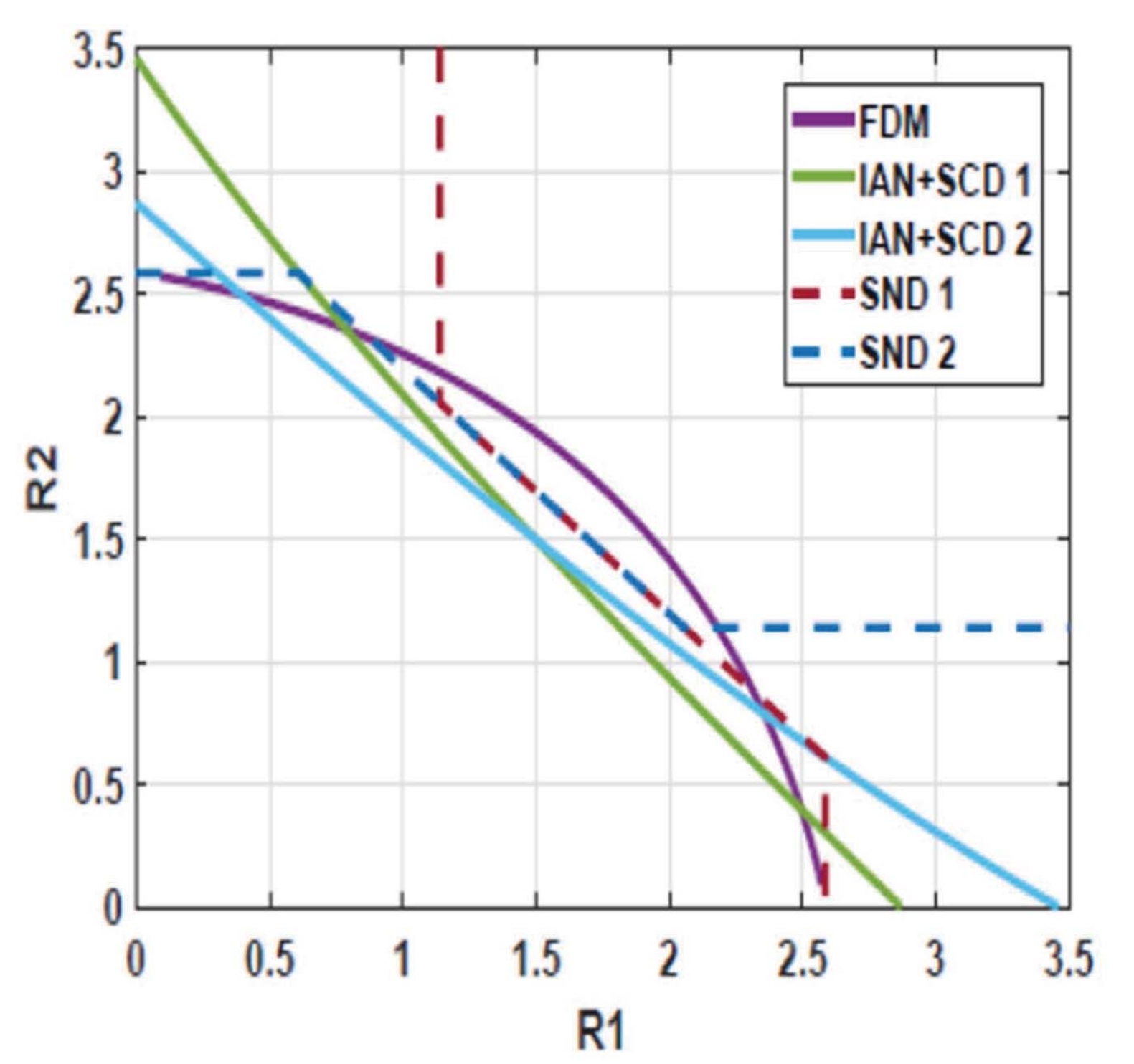}
       \centering
		\caption{Comparison of different rate regions.The power $P$ is varied to obtain the regions; $|h_{1}^{[1]}|^2=|h_{1}^{[1]}|^2=0dB$ and $|h_{2}^{[1]}|^2=|h_{1}^{[2]}|^2=-2 dB$.}
      \label{access}
\end{figure}

The authors in \cite{spaw} study these different strategies in order to go from frequency reuse 4 to 2; thus, improving spectral efficiency. As an example, Fig. \ref{access} compares the rates attained with some of the described strategies, in the specific case when the channels are unbalanced. Note that frequency division multiplexing (FDM) has also been included in the comparison in Fig.\ref{access} and it turns out to be sum rate optimal for a certain range of channel parameters within this class of computable achievable region. It is clear that the implementation of these strategies requires not only MUD capabilities of the UT, but also different transmission and resource optimization schemes (i.e., rate, power, bandwidth, time, beams). Also, different strategies and results are obtained.  Let us next give more details on the strategies for joint multi-user detection and management of resources.

\subsection{Joint Detection and Radio Resource Management}

Non-orthogonal access requires a new physical-layer, medium access control and resource allocation, which is an open topic in the V/HTS arena. Clearly, the way in which the different groups of users are clustered for scheduling, without rate splitting, affect the achievable data rate performance. In SatCom, the most representative scenarios differ from the terrestrial ones with respect to the traffic demand and the frequency reuse factor. This difference is basically due to the spatial-time correlation of the traffic in each beam, as each satellite beam has larger coverage area than a cellular terrestrial beam. In addition, the framing of the multi-user data that is used in the satellite protocol DVB-S2 is different from the one used in the terrestrial wireless standards. This is because satellite communication systems need larger channel coding gains than terrestrial counterparts. This different framing has important consequences in the user rate allocation. Therefore, these aspects motivate the need for different scheduling techniques. Also, when CSIT is available, it is also of interest to study how the high-performance MUD receivers can increase the data rate of precoding techniques in multibeam satellite systems. As an example, let us formulate the overloaded system:

\begin{equation}\label{genmud}
\begin{aligned}
& y_{k}^{[1]} = \mathbf{h}_k^{[1]H}\left(\mathbf{w}_k^{[1]} s_{k}^{[1]}+\mathbf{w}_k^{[2]} s_{k}^{[2]}\right)+z_{k}^{[1]}\\
& y_{k}^{[2]} = \mathbf{h}_k^{[2]H}\left(\mathbf{w}_k^{[1]} s_{k}^{[1]}+\mathbf{w}_k^{[2]} s_{k}^{[2]}\right)+z_{k}^{[2]},
\end{aligned}
\end{equation}
where two users per beam $k$ are considered. In addition, a different precoder $\mathbf{w}_k^{[i]}$ is allocated to each user $i=1,2$ in a beam; thus, the transmission is unicast. This differs from the multicast design of \eqref{sinr}, where the same precoder serves a group of users in a beam. As the beams share the same frequency and the number of feeds $N$ in the problem is less than the total number of users that are simultaneously served, $2N$, the system is overloaded; due to that the precoder is not able to eliminate all the interference. To overcome this issue, receivers use MUD techniques to deal with intra-beam interference. In this case, scheduling algorithms that are conceived for interference-free single-user detection techniques cannot be applied; instead, new algorithms to map users with beams are studied. Joint precoding and MUD can also be formulated and studied as previously done by the authors in \cite{globe}. The most convenient way to distribute the network resources in order to increase the access network capacity is still an open problem. As decentralized methods are needed to carry out this resource allocation in practice, many authors have presented promising recent developments based on concepts borrowed from game theory. Interesting examples of using game theory-based strategies applied to the future 5G wireless networks were presented in \cite{poor}. Coalition games, where sets of users form cooperative groups, can be seen as a suitable tool to study user clustering in multibeam HTS, as they are able to efficiently manage large-scale communication networks. Another alternative is matching game models, which can be used for decentralized user and sub-carrier pairing. All these are open topics for research.

The technologies presented in Section II and III have considered on-ground processing (either at the GW and/or UTs) to cope with the co-channel interference. The next section introduces the possibilities of onboard processing, that is, at the satellite.

%% file: obp.tex
The V/HTSs, once launched into Geostationary orbit, have a lifetime of about 12 to 15 years. This warrants including only the minimum necessary processing using viable technology that can support high bandwidths and sustain the constraints of satellite platforms, including power limitations, heat dissipation, and radiation. This has resulted in V/HTSs, which are typically based on link-budget design, being largely seen as passive relays performing only channelization and amplification on-board. Clearly, on-ground processing simplifies the payload architecture; in addition, such solutions are amenable to upgrades.

The advent of advanced processing like interference mitigation is being accommodated in such transparent satellite architectures through on-ground implementation. However, onboard processing (OBP), which provides for processing in the satellite, provides additional degrees of freedom that complement the on-ground processing (OGP). Particularly, these degrees of freedom can be used to enhance the following attributes:
\begin{itemize}
\item {\bf Latency}: Due to long round-trip delays, there is a large latency (250 ms) before the effect of OGP at one communication terminal is discovered at the other. The delay can be reduced by half through OBP, thereby enhancing the efficiency of the underlying techniques. This opens up the adaptation of OGP for onboard implementation, e.g., onboard precoding for mobile systems where round-trip delay affects fidelity of CSIT. 
\item {\bf Information Accessibility}: Since the satellite aggregates information from multiple GWs or UTs before appropriate channelization, it has more information than the constituent GWs or UTs. This enables joint processing on-board without the additional cost of sharing information across GWs/UTs, e.g., multiple GW joint processing.  
%
\item {\bf Support to Techniques}: Since many of the challenges and constraints arise onboard the satellite, OBP possesses the wherewithal to address them. OBP extends support for emerging techniques like full-duplex operations and anti-jamming techniques. For example, full duplex relaying by satellite requires OBP for canceling self-interference. 
\end{itemize}
This motivates an investigation into opportunities that emerge when the constraint of transparent satellites is relaxed. With most of the traffic carried over being digital (including TV), it is natural to consider onboard digital processing. 
\vspace*{-0.1in}
\subsection{State-of-the-Art in OBP}
Providing limited digital processing on-board the satellite is not a new concept and has been discussed in the last decades \cite{ref_10}. The key OBP paradigms observed from these developments can be categorized as follows:
\begin{itemize}
\item Regenerative processing is the straight-forward way to OBP; it involves generating the digital baseband data on board after waveform digitization, demodulation, and decoding. This is similar to the {\em decode and forward} paradigm in relay systems and is considered for multiplexing different streams, switching, and routing \cite{ref_9}. Clearly, regenerative processing provides better noise reduction and flexibility. However, its complexity is rather large for V/HTSs due to the high bandwidths used. Needless to say, such processing needs to be reconfigurable to accommodate evolutions in air-interface. 
%
\item A simpler approach to OBP is digital transparent processing (DTP), which operates only on the samples of the input waveform. The {\em amplify and forward} architecture in relaying is a simple DTP. Since neither demodulation nor decoding are implemented \cite{ref_10}, DTP processing results in payloads that are agnostic to air-interface evolutions. Typical applications include digital beamforming, broadcasting/multicasting based on single channel copies, RF sensing and path calibration \cite{ref_10}. 
\item An interesting hybrid processing paradigm involves not only digitizing the entire waveform, but regenerating only a part for exploitation. As a case in point, the header packet is regenerated to allow for onboard routing \cite{ref_12}. 
\end{itemize}
%

The OBP techniques used thus far have focused on networking such as onboard switching, traffic routing, and  multiplexing data/ multimedia \cite{ref_9}, with limited signal processing{\em per se}. However, as presented in the previous sections, a plethora of novel signal processing techniques has been considered of late for V/HTSs. These techniques would benefit from the additional degrees of freedom offered by OBP, hitherto not considered thereby motivating a study of onboard signal processing. The benefits of OBP are illustrated next through a simple signal processing application. 
%
%
\subsection{Interference Detection: Exploiting Different Flavors of OBP}
\label{ssec:Intf_Loc_Regen}
As an illustrative example, we consider the detection of interference at the satellite generated from on-ground terminals either maliciously or due to improper installation. These unwanted transmissions corrupt the desired signal being relayed, thereby reducing the end-user SINR and impacting the operations significantly. Currently, such interference is detected on-ground from downlink transmission and mitigated by operators using standard manual procedures. However, onboard interference detection can be undertaken by introducing a dedicated spectrum monitoring unit within the satellite payload that can take advantage of the emergent OBP capabilities. This provides for a faster reaction time and enhances detection capability; the latter arises due to avoidance of additional downlink noise and distortions from the satellite transponder, which affects on-ground detection \cite{Ch_ref4}.

We consider a generic system in which the satellite, the desired GW, and the interferer are equipped with one antenna. We further assume perfect digitization on-board.  Detection of the uplink RF interference can be formulated as the following binary hypothesis testing problem:
	\begin{eqnarray}
	& {\mathcal{H}_0}:& \widetilde{x}_k(n) = h s_k(n) + {\eta}_{1, k}(n), ~1\leq n\leq N,
	\nonumber \\
	&\mathcal{H}_{1}:&  \widetilde{x}_k(n) = h s_k(n) + {\eta}_{1, k}(n) + {p}_k(n), ~1\leq n\leq N,
	\label{eq:capacity}
	\end{eqnarray}
where $N$ is the number of samples, and $h$ denotes the scalar flat fading channel from the desired GW to the satellite. Further, let $s_k(n)$ be the sample of the intended signal transmitted by the desired GW on the $k$th channel  (or stream) at instance $n$; similarly, let ${p}_k(n)$ be the interfering signal on-board and  $\{{\eta}_{1, k}(n)\}$ be the noise modeled as a realization of independent and identically distributed (i.i.d.) complex Gaussian variables with zero mean and unit variance. 

\begin{figure}[t!]
	\centering
	\includegraphics[scale=0.45]{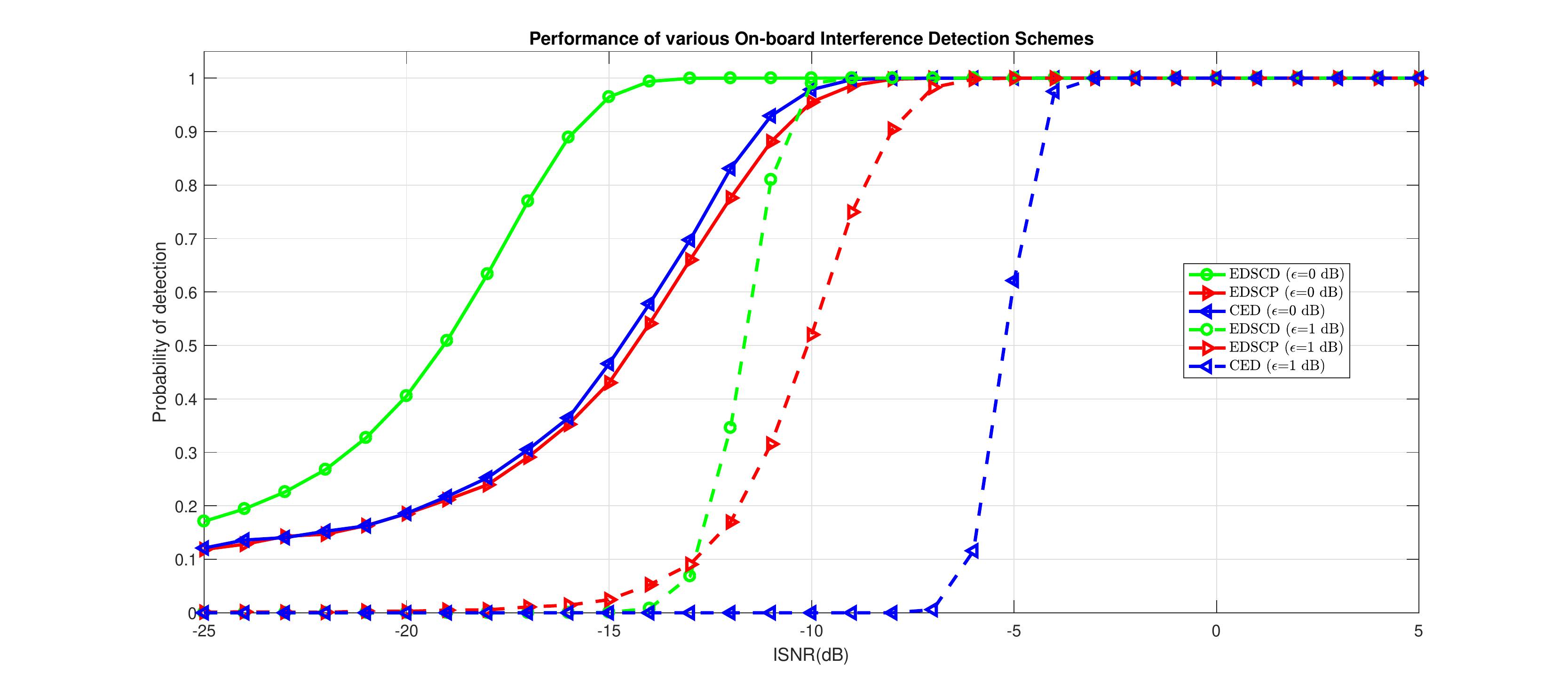} 
    \centering
	\caption{Probability of detection versus the SINR, QPSK modulation for $s_k(n)$, $N=516, SNR=6 dB$.}
		\label{fig:ID_res1}
	\vspace*{-0.2in}
\end{figure}

For such a problem, several interference detection techniques can be implemented on-board.
\begin{itemize}
\item The conventional energy detector (CED) technique works on samples $\widetilde{x}_k(n)$ directly  \cite{Ch_ref8}. CED is shown to be effective for strong interference and is susceptible to noise variations.
\item {\em ED with signal cancellation on pilots} (EDSCP) exploits the transmitted frame structure and estimates the channel, interference, $h, p_k(n)$, on pilot symbols (known $s_k(n)$) prior to energy thresholding \cite{Ch_ref10}.  This is an example of hybrid processing where only the frame header is decoded to ascertain the type of transmission and the location of pilots.  
\item   {\em ED with signal cancellation on data} (EDSCD), initially proposed in \cite{Ch_ref11} and further developed in \cite{Ch_ICASSP}, considers decoding of $\{s_k(n)\}$  and its subsequent removal, thus facilitating estimation of interference.  
\end{itemize}
Fig. \ref{fig:ID_res1} presents the probability of detection as a function of the received interference to signal and noise ratio (ISNR) comparing the following detection schemes: i) CED, ii) EDSCP, and iii) EDSCD. In practice, there is an uncertainty of 1 to 2 dB in the variance of ${\eta}_{1, k}(n)$ in \eqref{eq:capacity}; this uncertainty is represented as $\epsilon$ in the figure. We consider the number of modulated symbols and pilots as $N_d=460$, $N_p=56$, and $N=516$, representing a realistic waveform according to the DVB-RCS2 standard. It is observed that the interference detection performance decreases with uncertainty. The latter may lead to the ISNR wall phenomenon, where beyond a certain ISNR value the detectors cannot robustly detect the interference \cite{Ch_JSAC}. Furthermore, we see that the EDSCP and EDSCD schemes perform considerably better than CED with uncertainty, improving the ISNR wall by more than $5$ dB.
Thus, classical interference detection problems can be dealt with via different onboard architectures, with sophisticated processing providing additional performance benefits.  

Concerning interference, cancellation of narrowband interference in the RF chain is also an interesting topic. However, note that interference cancellation/ mitigation is typically preceded by an interference detection step. Thus, incorporation of such a process requires additional mass (analog components) or computation (digital components). Furthermore, the analog devices need to be adaptive as the nature of interference is unknown a priori. Keeping in mind the payload constraints and noting the development of on-ground procedures to turn off the localized interferer, interference cancellation is not considered in OBP. 


\subsection{System Model with an Onboard Processor}
\begin{figure}
	\centering
    \includegraphics[scale = 0.5]{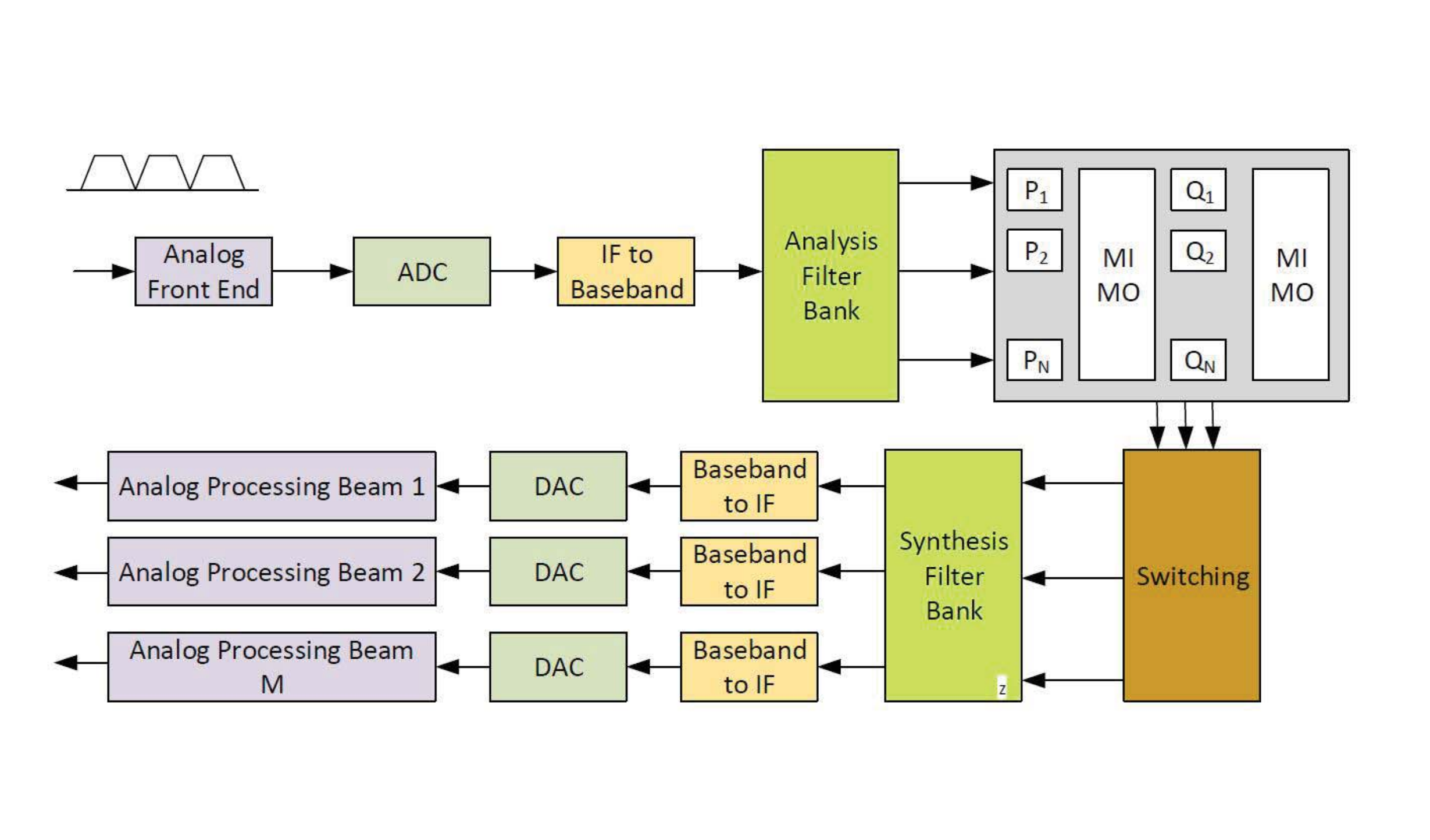}
    \centering
		\caption{Generic Architecture of Onboard Processor.}
      \label{fig:OBP_arch}
\end{figure}

%
\label{sssec:OBP_Arch}
Having demonstrated the usefulness of OBP with an illustrative example, we now proceed to detail the system involving OBP. Fig. \ref{fig:OBP_arch} presents a payload transponder employing digital OBP. Standard analog front-end receiver processing is carried out prior to the digital processing. These include filtering, low noise amplification and, mixer and automatic gain control; these are used in down-converting the input RF signal to an appropriate intermediate frequency (IF). 
%
The key components in OBP are detailed below: 
%
%
\paragraph{High-Speed Analog ADC and Baseband/IF Conversion} Assuming the IF signal with maximum bandwidth of $2f_c$ centered around $f_c$, ADCs sample at frequency $F_s \geq 4f_c$ to avoid aliasing. Subsequently, the resulting samples are converted to baseband (I/Q channels) using appropriate filtering \cite{ref:Sulli2}.
%
\paragraph{Analysis Filter Banks} The baseband/IF input is spectrally decomposed using a filter bank, where the output of each filter corresponds to the smallest quantum of user bandwidth. Typically, non-critically sampled implementation of the analysis filter bank is considered and a polyphase structure is used. Further, the filters are modulated versions of each other, leading to a fast Fourier transform (FFT) based polyphase matrix.
%
\paragraph{Processing Block} This generic block subsumes both transparent and regenerative architectures. It includes processing of individual streams (e.g., blocks $P_i, Q_i$) like demodulation, decoding as well as encoding, and modulation. The MIMO blocks impart joint processing capability. In the transparent architecture, these blocks can implement waveform manipulation techniques on one or more outputs of the filterbank; typical examples include a look-up table (LUT) for predistortion, beamforming, precoding, and spectrum calculation. 
%
\paragraph{Switching} The outputs of all transparent/regenerative processing chains are input to a switch matrix that effects routing in spatial (e.g., from one beam to another), temporal (e.g., store and forward), and spectral (e.g., frequency hopping) domains. The switching block is implemented through controlled memory reads and writes.
%
\paragraph{Synthesis Filter Bank and DAC} These implement the process of converting the digital samples in baseband to IF and finally to the RF domain; their implementation is similar to their counterparts $-$ ADC and analysis filter banks.
%
%

Critical to the implementation and operation of onboard processors is the accuracy of the processing chain. This is due to the limited link budgets of satellite systems. Hence, it is imperative to study various imperfections induced by the digital processing. These are listed below, and details can be obtained from \cite{ref:Sulli1}.
\begin{itemize}
\item Quantization errors induced by the ADC conversion
\item Non-idealities in filter implementation and use of fixed-point operations
\item Impairments due to phase noise and carrier offsets 
%
%
\end{itemize}
\subsubsection{Signal Model}
\label{sssec:Signal}
For ease of comprehension, we focus on a DTP here since the modelling of regenerative payloads is rather intractable. Let $x_k(t)$ be the analog signal corresponding to the $k$th frequency sub-band after the analysis bank. The signal $x_k(t)$ can be used to serve any beam after appropriate switching.  Assuming ideal filtering (i.e., rejection of out-of-band interference), a DTP would provide the designer access to the samples $\widetilde{x}(n)$ (at the input of the processing block in Fig. \ref{fig:OBP_arch}) where,
\begin{equation}
\label{eq:DTP_SM}
\widetilde{x}_k(n) \approx e^{j\left(2\pi [\triangle f] n + \theta + \omega_{n, k}\right)} x_k(nT_s +\tau_{n, k}) + \eta_{1, k}(n),
\end{equation}
where $[\triangle f]$ and $\theta$ are the frequency and phase offsets of the carrier assumed to be  independent of $k$, $\tau_{n, k}$ is the sampling jitter, $\omega_{n, k}$ considers phase noise, and $\eta_{1, k}(\cdot)$ represents the noise before the processing block. It should be noted that $\eta_{1, k}(n)$ possesses a flat spectral density over the sub-band $k$  only when the analysis filters are ideal. In addition, Doppler frequencies exist depending on the orbit; these can be included in $\triangle f$ as appropriate. 

Let $g_{D, k}(\cdot)$ be the functional equivalent of the OBP for the $k$th stream (after the analysis filtering without switching); to include the joint processing and synthesis filtering, $g_{D, k}(\cdot)$ can be typically modelled as a multiple input single output (MISO) function with memory. Further, let $g_{P, k}(\cdot)$ denote the equivalent transfer function of the payload for the $k$th sub-band. The output of the $k$th sub-band payload processing takes the form 
\begin{equation}
\label{eq:DTP_SM2}
y_k(t) = g_{P, k}\left(g_{D, k}(\left\{\widetilde{x}_l(m)\right\}_{l, m})+\eta_{2, k}(t) \right)+\eta_{3, k}(t),
\end{equation}
where $\eta_{2, k}(t)$ is the perturbation arising from the noise contributions from synthesis filter banks and DAC. Further, $\eta_{3, k}$ refers to the noise in the analog processing part of the $k$th sub-band. The presence of $\eta_{3, k}(t)$ in \eqref{eq:DTP_SM2} caters to the situation where the feeder link is non-ideal. 

It is clear from the earlier discussion that any algorithm operating on the digitized samples must be aware of:
\begin{itemize}
\item Perturbations in the input waveform as presented in \eqref{eq:DTP_SM}
\item Impact of processing on noise at the output of the payload as presented in \eqref{eq:DTP_SM2}
\end{itemize}
These make the algorithm design and its implementation challenging. In the ensuing section, an OBP algorithm design to minimize impairments will be discussed.
%
%
\vspace*{-0.1in}
\subsection{Impairment Cognizant on-board Predistortion using DTP}
\label{ssec:PRED_DTP}
As an illustrative example, we consider the implementation of predistortion on-board the satellite. This is motivated by the significant effort expended in the SatCom community on predistortion techniques towards countering the nonlinear impairments introduced by the onboard HPA and achieving higher power and spectral efficiencies \cite{BeSe10}. 

Signal predistortion (SPD), where the waveform is non-linearly transformed to mitigate the HPA non-linearities, has been considered in the literature (see references in \cite{Piazza_TSP} for a detailed list). These techniques have been traditionally implemented on-ground. By virtue of being a non-linear operation, SPD results in a bandwidth expansion;  when implemented on-ground, this can have serious consequences in SatCom systems where transmitted signals have to satisfy tight emission masks. Alternatively, additional constraints need to be imposed on SPD to satisfy the mask, which leads to diminished gains.  However, when using SPD onboard, the presence of output multiplexing (OMUX) naturally acts as a spectrum-regulating feature, thus relieving SPD design of additional constraints. Moreover, the LUT approach provides for an attractive alternative for implementation on-board rather than functional evaluation. 
\subsubsection{Modeling Onboard SPD}
\label{sssec:OBP_SPD}
\begin{figure}[h!]
	\centering
    \includegraphics[scale = 0.5]{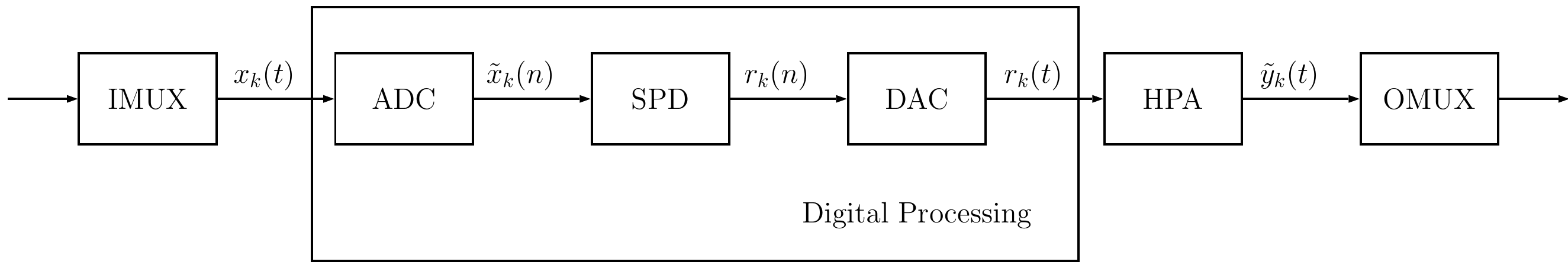}
    \centering
		\caption{Block scheme of the considered transponder with DTP.}
      \label{fig:block}
\end{figure}

We consider a simple payload model where the analog processing prior to DTP is denoted by an input multiplexing (IMUX) filter used to remove out-of-band signals from the signal being processed. Further, the analog processing subsequent to the DTP comprises HPA and an OMUX filter for mitigating out-of-band emissions. With reference to the DTP architecture in Fig. \ref{fig:OBP_arch}, we employ the processing blocks $\{P_k\}$, with block $P_k$ implementing SPD for the $k$th stream.  Focusing on a generic $k$th stream and omitting the filterbank for ease of analysis, the resulting payload architecture can be simplified to the schematic shown in Fig. \ref{fig:block}. 

In this setup, let $x_k(t)$ be the analog signal at the input of the ADC. An ideal ADC would perform the digitization, giving as output the sampled signal $x_k(n)=x_k(n/F_{s})$, where $F_{s}$ is the sampling frequency. However, this is seldom the practice; high-speed ADCs suffer from clock jitter caused by phase noise affecting the clock oscillator of the ADC. The jitter generates a non-uniform sampling of the input signal, leading to a degradation of the SNR inside the transponder \cite{Ra11}. Under some mild assumptions \cite{ToTiSa12}, the input of the SPD can be described as
\begin{equation}
\label{eq:SPD_jitter}
\widetilde{x}_k(n)	\cong x_k(n)+e_k(n)\dot{x}_k(n)+\eta_{1, k}(n)
\end{equation}
where $\dot{x}_k(n)$ is the sampled first-order derivative of $x_k(t)$ \cite{ToTiSa12} and $e_k(n)$ is the jitter error signal. The quantization noise and ambient thermal noise are included in $\eta_{1, k}(n)$.  A typical SPD model involves a third-order memoryless polynomial function taking the form 
\begin{equation}
\label{eq:SPD}
r_k(n)	=\gamma_k \widetilde{x}_k(n) +\delta_k \left|\widetilde{x}_k(n)\right|^{2} \widetilde{x}_k(n).
\end{equation}
On the other hand, the HPA can be modeled as a nonlinear non-invertible memoryless function approximated by a memoryless third-order Volterra series expansion as
\begin{equation}
\label{eq:twta}
\widetilde{y}_k(t) = \alpha_k r_k (t)+\beta_k \left|r_k(t)\right|^{2} r_k(t),
\end{equation}
where $r_k(n)$ is the sampled version of $r_k(t)$ using an appropriate sampling rate. Relating \eqref{eq:SPD}, \eqref{eq:twta} in the context of \eqref{eq:DTP_SM2} (ignoring filterbanks), we can immediately recognize that $g_{D, k}(\cdot)$ is a single-input single-output memoryless function taking the form $g_{D, k}(\widetilde{x}_k(n))= \gamma_k \widetilde{x}_k(n) +\delta_k \left|\widetilde{x}_k(n)\right|^{2} \widetilde{x}_k(n)$ and $g_{PL, k}(r_k (t))= \alpha_k r_k (t)+\beta_k \left|r_k(t)\right|^{2} r_k(t)$. Further, $\eta_{2, k}(t),\eta_{3, k}(t)$ are assumed to be zero. The aim is to determine the implementation gains of onboard SPD; central to such an implementation is the estimation of the predistorter coefficients minimizing a meaningful metric. This exercise is described below. 
\subsubsection{Parameter Estimation and SPD Implementation}
\label{sssec:OBP_PE}
As a first step, the HPA model parameters $\alpha_k, \beta_k$ need to be determined. They are jointly estimated by minimizing $\alpha,\beta	=\arg\!\min_{\mathbb{C}^{2}}E_{r_k}\left\{ \left\Vert \widetilde{y}_k(t)-y_k(t)\right\Vert ^{2}\right\}$, where $y_k(t)$ is the output of the actual HPA and $E_{r_k}\{.\}$ denotes the expectation with respect to the signal $r_k(t)$. This is a least-squares minimization problem, where the error function is linear in the coefficients $\alpha_k$ and $\beta_k$. We choose the least-squares cost function since (i) it represents the modelling errors and (ii) it is an elegant solution without the need for additional information.

The SPD parameters, $\gamma_k$ and $\delta_k$, can be found by resorting to a similar minimization problem, i.e., $\{\gamma_k,\delta_k\}	=\arg\!\min_{\mathbb{C}^{2}}E\left\{ \left\Vert y_k(t)-x_k(t)\right\Vert ^{2}\right\}$, where $E\{.\}$ denotes the expectation with respect to the signal $x_k(t)$, the jitter $e_k(n)$, and the noise $\eta_1(n)$.  Herein, we  again choose the least-squares cost function since it represents the MSE between transmitted and received symbols and does not require  additional information. However, now the error function is nonlinear in the coefficients $\gamma_k$ and $\delta_k$ because of  \eqref{eq:twta}, and the method used to estimate the Volterra coefficients of the HPA can no longer be used. Several methods have been proposed in the literature, including direct learning \cite{ZhDe07, Piazza_TSP}, which is based on least mean squares (LMS)/ recursive least squares (RLS), or divide and concur algorithm, which is based on message passing (see reference \cite{NM_SPAWC}).
%
%

While the SPD output $r_k(n)$ can be generated using \eqref{eq:SPD}, LUTs provide for a low complexity solution. Since the dynamics of the input signal, $\widetilde{x}_k(n)$ is typically known (from calibration tests for DTP), a LUT catering to this dynamic range can be calculated. Such a LUT implementation follows the memory-performance trade-off.
\subsubsection{Results}
\label{sssec:Re}
\begin{figure}[h!]
	\centering
    \includegraphics[scale = 0.65]{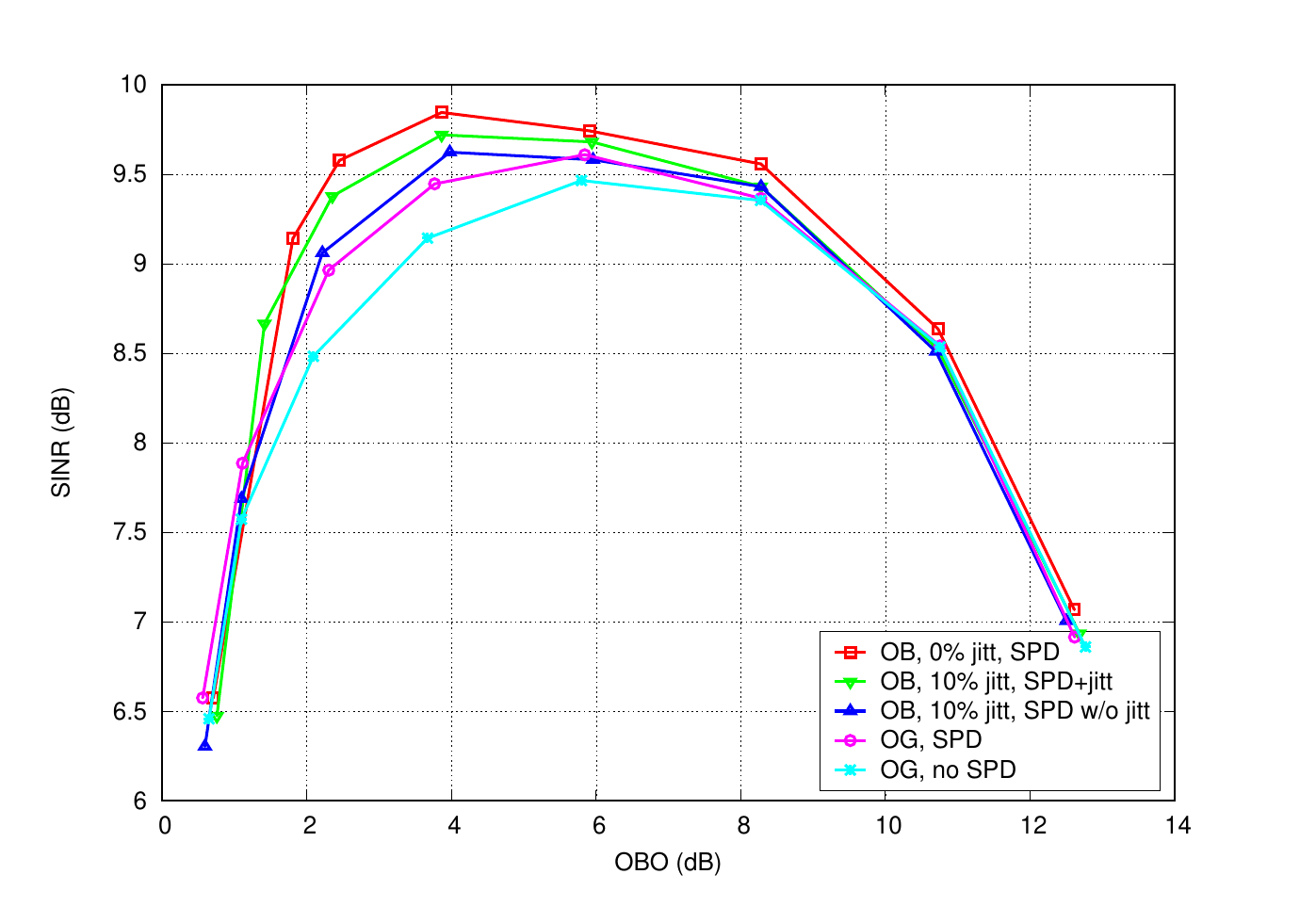}
    \centering
		\caption{SINR vs. OBO for the jitter cognizant onboard SPD, compared to its on-ground counterpart.}
      \label{fig:ground}
\end{figure}

Fig. \ref{fig:ground} compares the performance of the onboard SPD with its on-ground counterpart, where each has a similar structure \cite{NM_SPAWC}. The figure also illustrates the performance of onboard SPD optimized with and without the cognizance of jitter. Fig.  \ref{fig:ground} shows that the onboard SPD outperforms its on-ground counterpart both in terms of SINR (providing a 0.4 dB gain at output back-off, i.e. \textrm{OBO} = 4 dB) and in terms of OBO (providing a 2 dB gain at \textrm{SINR} = 9.5 dB). The on-ground SPD cannot compensate for the distortions of IMUX due to its memoryless nature, a shortcoming not encountered by OBP.  Fig.e \ref{fig:ground} also shows gains when jitter statistics are accounted for during the optimization of the SPD coefficients. To put the SINR gains in perspective, it should be noted that the DVB-S2X standard allows the use of a MODCOD with higher spectral efficiency \cite{DVB-S2X} for small SINR changes. Further, the OBO gain is of interest to satellite operators because it translates into a power-efficient amplification. 

This example shows how OBP provides a platform for signal processing algorithms to exploit, providing gains over OGP. This opens up avenues for investigation including sophisticated multi-stream interference mitigation techniques like precoding. 

%% file: flexible.tex
In the previous sections, we have focused particularly on physical-layer techniques that lead to increases in system throughput and user data rates. These techniques are essential in adapting the satellite systems to the increasing demands of the upcoming generation of communications networks such as 5G. However, additional gain can still be achieved by sharing spectra with other communication systems. This calls for cognitive satellite systems.

Another critical component of future communications networks is low latency. In this direction, caching techniques are proposed to deliver the content over the proverbial last mile, thereby reducing the latency significantly. Here satellite systems can help by provisioning multi/broadcast capabilities, and they can further help the terrestrial segment by off-loading traffic and reducing backhaul demand. 

Therefore, in this section, we discuss two flexible and hybrid terrestrial/satellite communication technologies that are particularly suitable for the 5G networks. In the first subsection, we discuss cognitive satellite systems and the underlying enabling techniques, followed by another subsection on integrated satellite-terrestrial backhauling for caching.

\subsection{Cognitive Satellite Communications}
\label{Cognitive Satellite Communications}
As mentioned earlier, the milestone for satellite communications is to reach 1 Tbps V/HTS satellite systems to cope with the expected demand in the upcoming generation of communications. To cope with such targets, on top of the signal processing techniques described before, additional bandwidth is necessary. However, considering the spectrum scarcity, the satellite networks will need to share the spectrum with the incumbent users to access the additional bandwidth.

In this direction, several scenarios have been studied in \cite{commag}, where describe the satellite systems sharing a spectrum with the incumbent satellite or terrestrial networks. In general, these scenarios can be divided into two categories: i) cognitive FSS downlink in the 18 GHz band shared with either microwave FS links or with broadcasting satellite services (BSS) feeder links, and ii) cognitive satellite uplink in the 28 GHz band shared with terrestrial microwave FS links.

The cognitive FSS downlink does not interfere with the incumbent users due to either sufficient satellite orbit separation or to limitations on the ground power flux density; rather, it is the cognitive terminal which receives the interference from the incumbent users. Therefore, in both cases, the cognition lies in mitigating the interference received from the incumbent users. On the contrary, in the case of cognitive FSS uplink, the cognitive terminal interferes with the incumbent users. In contrast to the FSS downlink, the uplink scenario becomes more similar to the conventional cognitive radio scenarios, in which the cognitive user has to protect the incumbent user from its interference. 

Irrespective of the fact that the cognitive terminal works in the downlink or the uplink, as in any other cognitive radio system, two enabling stages are considered: i) spectrum awareness, and ii) spectrum exploitation. Below, we discuss each of these techniques in detail and provide ideas for future work.

\subsubsection{Spectrum Awareness Techniques}
Spectrum awareness techniques can be categorized as spectrum sensing and radio environment mapping (REM) \cite{geert,rem}. Spectrum sensing refers to individual or cooperative cognitive radios that sense the spectrum and determine whether the incumbent user is available at a specific time, frequency, and geographical location, and if not, they use the spectrum. This area has been heavily studied over the last years, and several incumbent user-detection techniques such as energy detection, feature detection, and multiple antenna based techniques are proposed in the literature. 

On the other hand, REM considers a more general scenario, that is, producing a spectrum map of the environment either based on regulatory databases or on spectrum cartography techniques. Such a static or dynamic database can provide additional information with respect to spectrum sensing, which is only based on the detect-and-avoid mechanism. Examples of such information include the transmit power, channel gain, SINR, and interference level of the incumbent users. In this way, the spectrum can be exploited more optimally with respect to the sensing techniques.

Aside from the general advantages of REM, spectrum sensing has some serious limitations when applied to the cognitive satellite scenarios mentioned before. For example, in the case of cognitive uplink, detection of the incumbent user does not necessarily lead to protection of the incumbent receiver. As sensing the incumbent receiver is almost impossible in practice, spectrum sensing cannot be considered a reliable spectrum awareness technique in this case. 

Therefore, REM based on either regulatory databases or spectrum cartography seems the most promising spectrum awareness technique. A database usually contains information about the incumbent users, e.g., the transmit power, the location, altitude, type of antenna, direction of transmission, frequency band and carrier, polarization, and so on. Such information is usually sufficient to, for example, calculate the SINR of a specific user that would like to gain cognitive access to frequency band. While this is a more reliable and practical approach with respect to spectrum sensing, it has its own challenges. A database may not contain all the links, or the information about the links may not be complete. Further, a database needs regular updating as the information about the incumbent users may become outdated. Therefore, dynamic database building techniques are suggested such as spectrum cartography to either produce a database or update and complete the existing database. 

In contrast to databases, spectrum cartography techniques mostly rely on cooperative spatial-temporal estimation techniques to recover information about the incumbent users. While several successful attempts are described in the literature to develop such techniques for omni-directional incumbent users, considering the highly directional nature of the users in higher parts of the spectrum, e.g., the Ka band, specific techniques considering such directivity issues need to be developed. Works such as \cite{dot} have made some initial steps in this direction. However, application of more advanced signal processing techniques such as statistical learning, as in \cite{gonzalo}, should be considered in order to develop more efficient and reliable results.

In the case of cognitive satellite communications, the spectrum awareness leads to production of a SINR matrix of the users in each available carrier as follows \cite{eva}:

\begin{equation}
	\textbf{SINR}= \begin{bmatrix}  \text{SINR}(1,1) & \cdots & \text{SINR}(1,K) \\ \vdots & \ddots & \vdots \\ \text{SINR}(M,1) & \cdots & \text{SINR}(M,K)                    \end{bmatrix},
\end{equation}
where $M$ is the number of carriers and $K$ is the total number of FSS terminals. Each element of $\textbf{SINR}$ is calculated by
\begin{equation}
	\text{SINR}(m,k)=\frac{P(k)}{I_k(m)+I_{\text{co}}+N_0},
\end{equation}
where $I_k{(m)}$ is the aggregated interference received at the FSS terminal from the FS stations in the FSS downlink scenario or the imposed interference on the FS terminals in the FSS uplink scenario. Furthermore, $I_{co}$ is the co-channel interference coming from the multibeam nature of the satellite system, and $P(k)$ is the received power in the case of FSS downlink, or the maximum allowed transmit power in the case of FSS uplink. As mentioned earlier, this information can be obtained either from a database or by measurement at the FSS terminals in the FSS downlink scenario or by spectrum cartography in the FSS uplink scenario. Furthermore, techniques such as receive beamforming in the case of cognitive satellite downlink or transmit beamforming in the case of cognitive satellite uplink as in \cite{eva} can be applied to reduce the received interference in the cognitive satellite downlink or reduce the imposed interference in the cognitive satellite uplink. For details on how to derive the underlying parameters, e.g., $P(k)$, and $I_k(m)$, we refer to \cite{eva}. 

Having this information, in the next subsection we discuss current spectrum exploitation techniques.

\subsubsection{Spectrum Exploitation Techniques}
Having the knowledge of the spectrum, the next step is to allocate the resources accordingly. To do so, several different types of optimization problems can be defined depending on the objectives of the designer. Here, we focus on the most common one of interest to the satellite industry, that is, maximizing the system throughput or sum-rate. Works such as \cite{evaicc} consider other objective functions as well, such as multi-objective, max-min, and proportional fairness. Furthermore, in this paper, we only focus on the case a single satellite operator; extension of cognitive access to the case of multiple operators with limited information exchange is a challenge with potential for future studies.

Based on the received $\textbf{SINR}$ from the spectrum awareness stage, the underlying maximin sum-rate problem can be formulated as follows,
\begin{equation}
			\max_{\textbf{A}}\ \ \  \left\| \text{vec}(\textbf{A} \odot \textbf{R}(\textbf{SINR})) \right\|_{l_1} \ \ \ \ \ \			\text{s.t.}\ \ \   \sum^{K}_{k=1} \textbf{a}_k(m) = 1,
	\label{ca_eq}
\end{equation}
where ${\textbf{A}}$ is the carrier allocation matrix with its elements ${\textbf{A}}(m,k)=\{0,1\}$. The constraint $\textbf{a}_k(m) = 1$ takes care of the orthogonality of the carrier assignment imposed by the SatCom standards, e.g., DVB-S2 or DVB-RCS2. \cite{eva} shows that this problem can be solved by applying the Hungarian algorithm in polynomial time. Further, it is shown that the above problem can lead to up 600$\%$ improvement in satellite downlink throughput, as well as 400$\%$ improvement in satellite uplink throughput. 

While problems such as \eqref{ca_eq} provide an efficient method of allocating resources for cognitive satellite systems, there are still several challenges to be addressed by the signal processing community. First, in the case of multiple satellite operators, particularly for the FSS uplink scenario, the operators needs some level of coordination among themselves to make sure the aggregated interference does not go beyond a specific threshold. In this example, the SINRs of individual users depend not only on their own transmit power, for example, but also that of other users. This calls for distributed resource allocation mechanisms. 

Furthermore, the current approaches mostly address the orthogonal carrier assignment; application of NOMA could lead to higher system throughputs and also uncover further challenges to be solved by academic and industrial researchers. Finally, the current techniques focus mostly on power and carrier allocation, while bandwidth is another flexible parameter that can be assigned to users. Joint optimization of power, carrier, and bandwidth is another open problem in this direction.

Another similar problem, which has been recently studied, is the resource allocation for integrated satellite-terrestrial systems. In such problems, both satellite and terrestrial systems enjoy the same access priority. Solving this problem is particularly motivated by the prospect of 5G backhauling. In \cite{sansa2}, a problem in this direction is defined by
\begin{equation}
{\text{max}_{P_{t},P_{s},\bf{A}_{t},\bf{A}_{s}, \textbf{BW}_{t},\textbf{BW}_{s}}}  \sum^{L}_{j=1}R_{t}(j)+\sum^{M}_{j=1}R_{s}(j),
\end{equation}
where $R_{t}$ and $R_{s}$ are the respective terrestrial and satellite stations' throughput, with $L$ and $M$ being the total number of terrestrial and satellite stations, respectively. Furthermore, $\bf{P}_{t},\bf{P}_{s},\bf{A}_{t},\bf{A}_{s}, \textbf{BW}_{t},\textbf{BW}_{s}$ are the respective transmit power of the satellite and terrestrial systems, the carrier allocation for the terrestrial and satellite systems, and finally the carrier bandwidth of the terrestrial and satellite systems. Depending on the specific scenarios and possibilities, such as NOMA, beamforming, and so on, different constraints can also be defined for this problem. Such problems are in general non-convex and NP hard, and thus some relaxation or heuristic techniques are necessary to solve the underlying problems. In \cite{sansa2}, an alternating descent approach is adopted to solve a specific version of this problem with the goal of assigning only the carriers, assuming the power and bandwidth of the two systems are known. Nevertheless, the convergence of the algorithm in this case is not guaranteed. Therefore, resource allocation techniques and efficient solutions for this and other possible problem are other open avenues to the signal processing community for future research.

\subsection{Integrated Satellite-Terrestrial Backhauling Architectures for Caching}

In the context of 5G communication systems, satellite backhauling has emerged as one of the most prominent use cases towards satellite-terrestrial system integration. The main reason behind this is the capability of V/HTS to sustain fat pipes towards designated areas where terrestrial infrastructure is insufficient or entirely absent. 
Current state-of-the-art systems are targeting transponder bandwidths up to 2 GHz, which in combination with highly efficient MODCODS, can deliver rates of more than 10 Gbps. 
The most important challenge for future networks has been widely recognized to be mobile video in the form of video streaming. According to a CISCO white paper \cite{intro_Cisco_VNI}, 
“mobile video will increase 11-fold between 2015 and 2020, accounting for 75\% of total mobile data traffic by 2020”. As unbelievable as this percentage may seem, it comes into perspective when considering the evolution of consumer trends related to video content. With the proliferation of smart devices, more and more users are switching from the traditional linear broadcasting services (TV channels) to streaming services, such as YouTube, NetFlix etc. Another factor that contributes to the traffic projections is the increasing video quality, i.e. 3D, 4K video, virtual reality etc., which can be translated to increased bandwidth requirements for both the core and access networks.  
To address this challenge, major telecom operators have been building their own content delivery networks (also known as telco CDNs).

\begin{figure}[h!]
	\centering
    \includegraphics[width=0.5\columnwidth]{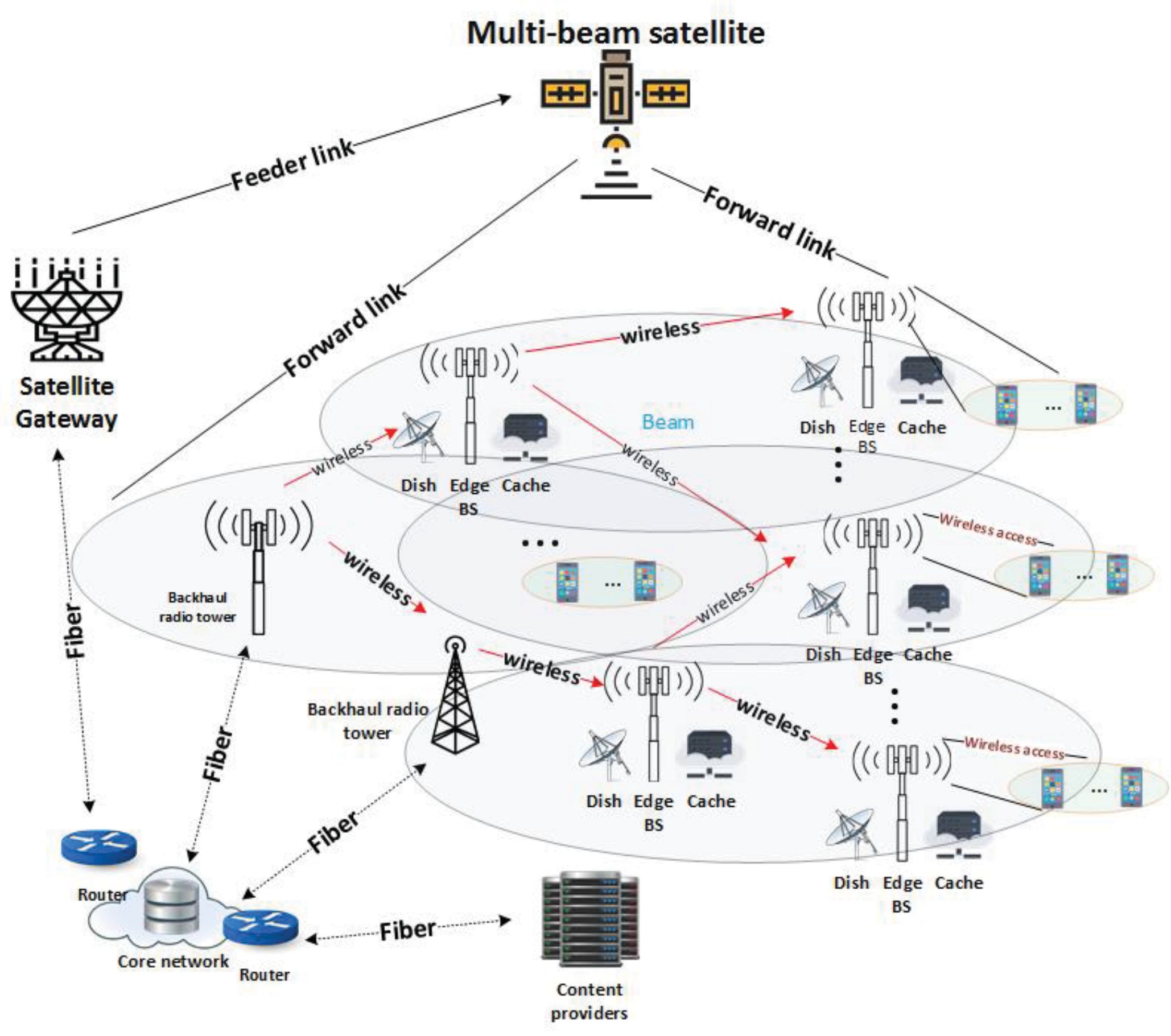}
    \centering
		\caption{Integrated satellite-terrestrial architecture for caching.}
      \label{caching}
\end{figure}

In this direction, integrated satellite/terrestrial architectures can be used to optimize the backhaul capacity needed for caching. This is a key element since satellite networks can deliver large amounts of delay-tolerant information with just two hops to multiple edge nodes through geocasting (broad/multicasting across large geographic areas). This will take the load off the multihop terrestrial backhaul networks, which can then be used to deliver the unicast broadband services that they were originally designed for.
Such an integrated satellite-terrestrial architecture for a telco CDN is depicted in Fig. \ref{caching}. Compared to traditional CDNs (e.g., Akamai), this architecture can accommodate deeper edge caching (e.g., at the base station), as it controls the last mile. Furthermore, it can control the full protocol stack (from PHY to application layer), allowing for improved cross-layer optimization, as well as better profit margins. Focusing on the technical advantages, traditional CDNs usually base their decisions on limited information about the underlying infrastructure. However, integrated telco CDNs have full control of the infrastructure in terms of understanding bottlenecks (e.g., peak hours, failed links) and utilizing suitable transmission modes (uni/multi/broadcasting). In this direction, telcos can optimize the delivery of content to the edge caches in order to minimize the required resources without interfering with the normal operation of the network.

As it can be seen in Fig. \ref{caching}, we focus here on a V/HTS multibeam satellite in accordance with the previous sections. This multibeam configuration might operate using aggressive frequency reuse or spectrum coexistence as described in Sections \ref{Precoding in Multibeam Satellite Systems} and \ref{Cognitive Satellite Communications}, respectively. Each edge base station (BS) is equipped with a limited memory cache and two interfaces to receive the content, one for terrestrial backhauling and the other for satellite backhauling. In these architectures, both satellite and terrestrial networks are used to place the files in the caches of each cache-enabled edge cellular BS. The main advantage is that the satellite can deliver the cached content using multimodal backhauling. More specifically, due to the beam pattern structure, the transmission can support various granularities. The broadcast mode can support transmission towards the entire coverage area (e.g., entire continent). In the multicast mode, it is localized towards a cluster of beams (large country or equivalent for linguistic beams) or even towards a single beam (large metropolitan area or small country). At the same time, the terrestrial connectivity can allow a unicast mode where individual BSs can be targeted. This level of flexibility allows for highly efficient content distribution, provided that the system has side information about the content popularity across the coverage area. 
In this context, the work in \cite{Kalantari2017} has focused on off-line caching and has shown that this architecture can improve the cache hit ratio while reducing the overall required backhaul rate. In a more general context, the application of satellite communications in feeding several network caches at the same time using broad/multi-cast has also been investigated in \cite{Linder2000}, \cite{Brinton2013}. The work of \cite{Linder2000} proposes using the broad/multi-cast ability of the satellite to send the requested content to the caches located at the user side. Online satellite-assisted caching is studied in \cite{Brinton2013}. In this work, satellite broadcast is used to help place the files in the caches located in the proxy servers. Each server uses the local and global file popularity to update the cache. Pushing content to the caches using hybrid satellite-terrestrial networks is investigated in \cite{Evans2015}.

\subsubsection{Resource Allocation for Multimodal Content Delivery}
One of the most important problems in integrated satellite-terrestrial systems is the resource allocation across the multiple modes of delivery. More specifically, each mode has a different utilization cost and can deliver a different amount of useful bits to the user population. For example, unicast transmission through the terrestrial networks is usually the most cost-effective but it can deliver information only to a specific location/user. On the other hand, broadcast is less cost-effective, but the delivered number of bits is multiplied by the number of users reached. This trade-off poses interesting challenges in multimodal content delivery, as explained in the following paragraph.

First, let us define a popularity measure based on the number of requests, which is tractably described though a function. A widely used abstraction for this function is the Zipf law, which is given by the pmf
$
f[i] =  F^{-1}\left(\frac{1}{i}\right)^a\label{eq: Zipf}
$, 
where $ F = \sum_{i=1}^{I}\left(\frac{1}{i}\right)^a $ is the total number of recorded requests for a library of $I$ files.
In more detail, if we ordered the files from most to least popular, then the relationship governing the frequency at which the file of rank $i$ will appear is given by \eqref{eq: Zipf}. Consequently, the probability of a request occurring for file $i$ is inversely proportional to its rank, with a shaping parameter $\alpha$.  

For the sake of simplicity, let us assume here only two modes: the satellite broadcast mode and the terrestrial unicast mode. The resource allocation problem focused on defining a popularity threshold above which multimedia content is broadcasted and below which it is unicasted. Each of the $K$ BSs requires only one file of size $s$ [bits] out of \(I\) files, thus $F=K$.  Let $\hat i$ denote the optimal threshold and $R_\mathrm{bc}$ and $R_\mathrm{uc}$ denote the supported rates of the broadcast and unicast modes, respectively. Then the transmitted volume of data in BC mode will be given by $ V_\mathrm{bc}= (\hat i -1)\cdot s $, since each file has to be broadcasted only once. The received/cached data volume through the BC mode across all \(K \) users will be $ s \cdot  K \sum_{i=1}^{\hat i-1}f[i]$, where the last coefficient determines the percentage of BSs requesting the broadcasted files. On the other hand, in unicast mode, the transmitted volume of data will be  $V_\mathrm{uc}=s\cdot K  \cdot\sum_{i=\hat i}^{I}f[i]$,  since each file is individually transmitted to each user that has requested it. As a result,  the received volume is by definition equal to the transmitted volume. The goal here is to find the optimal threshold $\hat i$ that minimizes the total transmission time\footnote{It should be noted that an equivalent formulation can be expressed in terms of fixing the required time and minimizing the required bandwidth.} by formulating the problem below:
$\mathcal{T:}\  \hat i=\arg\min_{i}  \ T_\mathrm{tot} = T_\mathrm{uc} +T_\mathrm{bc}\label{eq: cost}
$,
  where $T_\mathrm{uc} = V_\mathrm{uc}/R_\mathrm{uc}$ and $T_\mathrm{bc} = V_\mathrm{bc}/R_\mathrm{bc}$. For $\alpha > 1$ in \eqref{eq: Zipf}, the optimization problem  \eqref{eq: cost} can be straightforwardly solved by setting the function’s derivative equal to zero. The optimal threshold between broadcasting and unicasting modes is given by optimizing the following cost function
  
\begin{equation}
T_{tot}(\hat i) =  s\cdot\left(K\frac{\sum_{i=\hat i}^{I}f[i]}{R_{\mathrm{uc}}}+\frac{\hat i-1}{R_{\mathrm{bc}}}\right).
\end{equation}

\begin{figure}
\centering
\includegraphics[width=0.5\columnwidth]{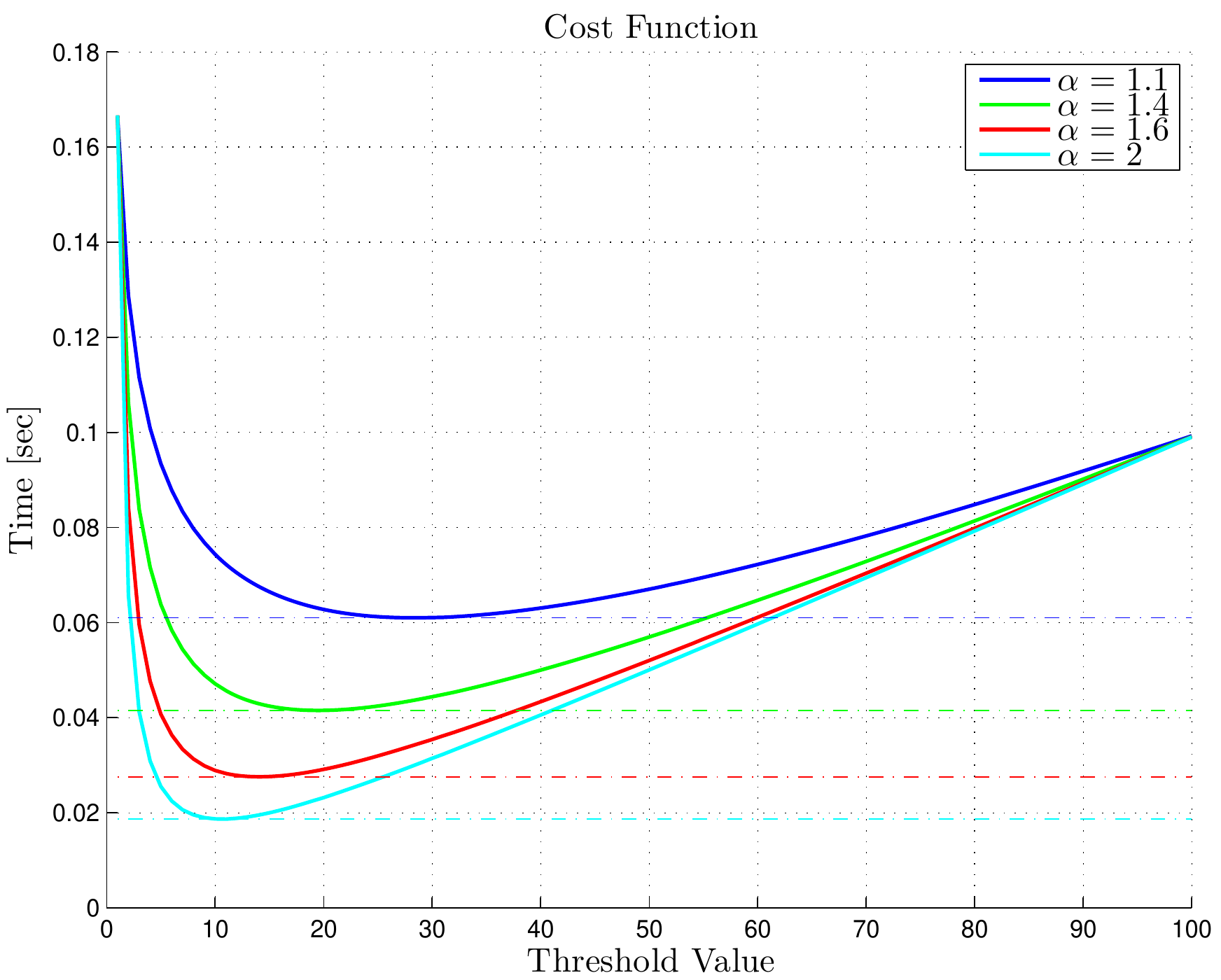}\\
\centering
\caption{Cost function versus a varying threshold value \(\hat i\), for various popularity function parameters \(\alpha\). The minimum value of each function is also pointed out. }\label{fig:  cost}
\end{figure}

Assuming \(K=500\) BSs, a database of \(I=100\) files and a rate ratio of $R_\mathrm{uc}/ R_\mathrm{bc}=3$, the cost function of problem $\mathcal T$ is plotted in Fig. \ref{fig:  cost} for a range of parameters. It should be noted that the two extremes of this figure represent the time needed by unicast only (far left) and broadcast only (far right). 

In V/HTS systems the actual rate ratio between unicast and broadcast depends on many parameters such as the system architecture and the employed transmission schemes. One option would be to incorporate the broadcast service as a separate payload in the V/HTS mission or even as a separate satellite. Alternatively, it could be accommodated through the multibeam architecture by occupying part of the V/HTS transponders. Similarly, the unicast service could operate with conventional 4-color frequency reuse or with aggressive full-frequency reuse through precoding. All the aforementioned design decisions will affect the rate ratio and as a result the optimal loads balance between the two modes. 

\subsubsection{Coded Caching}
In the previous section, we have focused only on uncoded caching, where the content is stored at the BSs without any preprocessing or coordination. Recently, the paradigm of coded caching was proposed, where the content is preprocessed before being transmitted. More specifically, during the content placement phase, a different combination (XOR) of files is designed and transmitted individually towards each BS. This way, an additional caching gain can be achieved during the delivery phase by broadcasting a combination of files. Each broadcast is received by multiple BS, allowing them to obtain simultaneously different parts of the files based on the information stored during the placement phase. It is shown in \cite{CachingLimits} that the coded caching achieves a global caching gain on top of the local caching gain. This gain is inversely proportional to the total cache memory. At a first glance, this method seems suitable for V/HTS, as the broadcast mode is inherently available in order to facilitate the delivery phase. However, there is a price to be paid, namely the overhead during the placement phase, where strictly unicast transmission is needed. This can be easily perceived taking into account that each BS has to store a different preprocessed XOR combination of files. In other words, in the context of integrated satellite-terrestrial architectures, the placement phase of the coded caching approach should be handled by the terrestrial segment. Another critical aspect of coded caching is that every time the file database changes, a network-wide cache update is needed to support the newly added files. This process is much easier with conventional caching as only part of the BS population or BS cache has to be updated.

%% file: conclusions.tex
V/HTSs offer far more resources than their ancestors. This article covered the main signal processing challenges and tools that can be used to boost the spectral efficiency that these satellites can offer. We first discussed high frequency reuse in the multiple beams of these satellites and the role of on-ground processing (OGP), both at the gateway and at the user terminal: precoding at the gateway site and advanced multiuser detection at the user terminal. Next, with most of the traffic carried over being digital, signal processing onboard the satellite (OBP) comes as a natural consequence; this overcomes some drawbacks of OGP. Finally, additional gains can be achieved by sharing the spectrum with other communications systems. This motivated the last section on flexible communications and hybrid satellite-terrestrial solutions, which aim at a seamless integration of satellites into future 5G networks to ensure ubiquitous coverage. 

The focus of the article has been FSS. However, SatCom refers to a wide range of systems operating in various frequencies and providing different types of services. Future prospects for this work include the application and adaption of the explained signal processing techniques to the MSS. There are also high expectations in the so-called mega-constellations of MEO and LEO satellites. In these cases, the signal processing concerns are Doppler compensation, and inter-satellite communications, to mention just two. They require scanning user terminals that are more complex than the GEO ones. The feasibility of such solutions depends very much not only on their complexity level but also on their compatibility with the legacy systems, full integration with the wireless terrestrial communications networks, user terminal complexity, and policy.